\PassOptionsToPackage{table,xcdraw,svgnames}{xcolor}
\documentclass{interact}

\usepackage[english]{babel}

\usepackage{parskip}
\usepackage[onehalfspacing]{setspace}

\usepackage[stretch=10]{microtype}

\usepackage{csquotes}

\usepackage{graphicx}
\usepackage{tikz}
\usetikzlibrary{trees}
\usepackage{wrapfig}

\usepackage{booktabs}
\usepackage{tabularx}
\usepackage{tabulary}
\usepackage{multirow}
\usepackage{rotating}
\usepackage{longtable}
\newcolumntype{s}{>{\setbox0=\hbox\bgroup}c<{\egroup}@{}}
\usepackage{algorithm}
\usepackage{algpseudocode}

\usepackage{amsmath, amssymb, amsfonts, bbold}

\usepackage{listings}

\usepackage{subcaption}

\usepackage{natbib}
\bibpunct[, ]{(}{)}{;}{a}{}{,}
\renewcommand\bibfont{\fontsize{10}{12}\selectfont}

\definecolor{navy}{RGB}{0, 0, 128}

\usepackage{float}

\usepackage{enumitem}

\usepackage[colorlinks=true,
            linkcolor=navy,
            citecolor=navy,
            urlcolor=black,
            filecolor=black,
            pdfborder={0 0 0}]{hyperref}

\usepackage{credits}

\credit{AR} {1,1,1,0,1,1,0,1,1,0,1,1,1,1}
\credit{JS} {1,0,0,0,0,1,1,1,0,1,0,1,1,1}
\credit{FD} {0,0,0,0,0,1,0,0,0,0,0,0,0,1}
\credit{MP} {1,0,0,1,0,1,0,0,0,1,0,0,1,1}

\usepackage{ifthen}

\newboolean{anonymous}
\setboolean{anonymous}{false} 

\begin{document}


\title{Which Imputation Fits Which Feature Selection Method? \\ A Survey-Based Simulation Study
}

\ifthenelse{\boolean{anonymous}}{
    \author{}
    \date{}
}{

\author{
\name{Andrés Romero\textsuperscript{a,*}, Jakob Schwerter\textsuperscript{a,*}\thanks{CONTACT Jakob Schwerter (\href{https://orcid.org/0000-0001-5818-2431}{ORCID: 0000-0001-5818-2431}), TU Dortmund University, Department of Statistics, Martin-Schmeisser-Weg 13, D-44227 Dortmund, Germany. Email: \texttt{jakob.schwerter@tu-dortmund.de}.}, Florian Dumpert\textsuperscript{b} and Markus Pauly\textsuperscript{a,c}}
\affil{\textsuperscript{*}Shared first authorship;\\
\textsuperscript{a}TU Dortmund University, Dortmund, Germany;\\ \textsuperscript{b}Federal Statistical Office of Germany, Wiesbaden, Germany;\\
\textsuperscript{c}Research Center Trustworthy Data Science and Security, Dortmund, Germany}\thanks{%
    The project ``From Prediction to Agile Interventions in the Social Sciences (FAIR)'' is receiving funding from the program ``Profilbildung 2020'', an initiative of the Ministry of Culture and Science of the State of Northrhine Westphalia. The sole responsibility for the content of this publication lies with the authors.\\
    This study was not preregistered.\\
    The data is based on \cite[][SC6:13.0.0]{neps_network_2022} and can be requested at www.neps-data.de/Mainpage.
  }
}
}



\maketitle
\begin{abstract}
Tree-based learning methods such as Random Forest
and XGBoost are still the gold-standard prediction methods for tabular data.  Feature importance measures are usually considered for feature selection as well as to assess the effect of features on the outcome variables in the model. This also applies to survey data, which are frequently encountered in the social sciences and official statistics. These types of datasets often present the challenge of missing values. The typical solution is to impute the missing data before applying the learning method. However, given the large number of possible imputation methods available, the question arises as to which should be chosen to achieve the 'best' reflection of feature importance and feature selection in subsequent analyses. In the present paper, we investigate this question in a survey-based simulation study for eight state-of-the art imputation methods and three learners. The imputation methods comprise listwise deletion, three MICE options, four \texttt{missRanger} options as well as the recently proposed mixGBoost imputation approach. As learners, we consider the two most common tree-based methods, Random Forest and XGBoost, and an interpretable linear model with regularization.
\end{abstract}


\section{Introduction} 
\label{sec:Intro}

Machine learning (ML) is increasingly crucial for all kinds of quantitative analyses. This particularly holds for official statistics \citep{beck2018machine, unece2021mlinofficialstats, dumpert2023mlingerman, measure2023sixyears, dumpert2024maschinelles} based on survey data, which is of particular importance in countries such as Canada or Germany. Data processing includes integrating data from different sources, classifying and coding units, reviewing and validating input data (e.g., validating data against pre-defined edit rules), as well as editing and imputing incorrect, missing, unreliable or outdated information \citep{unece2019gsbpm}. Social science research presents a similar case. Here, ML offers significant advantages in areas like double machine learning for selecting key control variables \citep{ChernozhukovEA2018, ChernozhukovEA2024} and predicting student success with learning analytics. Unlike traditional regression methods, which excel at confirmatory hypothesis testing but limit the number of predictor variables \citep{ByrnesM2007}, interpretable machine learning techniques incorporate built-in feature importance and selection. Such capabilities are critical in contexts such as large-scale assessment and learning analytics, where the potential pool of control variables is large \citep{ArizmendiEA2022, BelloniCH2013, SchwerterEA2024}.

Complex machine learning models, such as deep neural networks, often sacrifice interpretability, a drawback that has led to the development of interpretable machine learning \citep{molnar2020interpretable, ewald2024}, which is now widely used in the social sciences \citep{foster2020bigdata}. When discussing quality aspects of machine learning in official statistics \citep[p. 200, 202]{vandelden2023tenprop}, it is advisable not to give too little thought to feature selection, manual or automatic, see also \citet{molladavoudi2023exploring} and \citet[Section 3.3]{saidani2023qualitatsdimensionen}. Approaches including Lasso, Elastic Net regression, and tree-based models such as Random Forest and XGBoost are widely used for high-dimensional data and feature selection, see, e.g., \citet{ParkEA2023}, \citet{LupyanG2019}, and \citet{ArizmendiEA2022} for some examples in learning analytics. These methods can handle large sets of variables while preserving interpretability -- an essential factor in social science research and, albeit to a somewhat lesser degree, in official statistics.

Missing data poses a significant challenge in the application of machine learning, particularly in psychological and social science research. For example, \citet{peng2006advances} found that at least 48\% of articles in eleven education journals between 1998 and 2004 reported missing data, highlighting the prevalence of this issue. Ignoring missing data can lead to bias, reduced statistical power, and increased standard errors \citep{vanBuuren2018, dong2013principled}. Recognizing this, advanced techniques like multiple imputation were introduced by \citet{schafer2002missing}, building on \citet{rubin1976inference} and further developed by \citet{rubin1987multiple}. According to \citet{enders2023missing}, this work is the most cited paper in Psychological Methods, underscoring its impact. Recent technological and software advancements have improved imputation models \citep{enders2023missing}, facilitating their adoption in the social sciences and official statistics, where imputation is also common practice \citep{chen2019recent}. This underscores the necessity to evaluate whether machine learning methods can effectively handle missingness, thereby enhancing their applicability in fields where missing data is prevalent.

While multiple imputation effectively handles missing data, researchers have to choose from several imputation methods \citep{dagdoug2023impproc, ElBadisyEA2024, schwerter2024evaluating}. Among these, Multiple Imputation by Chained Equations (MICE) is very popular, with Predictive Mean Matching (PMM) being one of the most widely used strategies \citep{VanBuurenGO2011, vanBuuren2018}. MICE with PMM is broadly used in social science research because of its flexibility in handling different patterns of missing data and taking into account dependencies among variables \citep[e.g.,][]{GopalakrishnaEA2022, VanGinkelEA2020, ZettlerEA2022, HajovskyCJ2020, HollenbachEA2021, WestermeierG2016, DeFranzaEA2021, CostantiniEA2023}. The fact that theoretical or empirical relationships exist between variables is also often used in official statistics to define so-called edit rules. Responses and imputations must fulfill these edit rules in order to be considered plausible. Correctly mapping the dependencies between variables is therefore often an important additional criterion for a good imputation \citep[p. 299]{Waal.2011}. 

There are also alternative MICE methods that have good properties and are in widespread use. For example, MICE Norm \citep{VanBuurenGO2011, vanBuuren2018} uses a Bayesian linear normal model to predict missing values, and recent simulation studies in official statistics favor MICE Norm for its high univariate and multivariate imputation accuracy with metric variables \citep{thurow2021imputing, thurow2024assessing}. This approach effectively preserves the underlying distributions and accurately reflects true correlations. In addition, biostatistical research has shown that MICE Norm provides reliable type I error control after imputation \citep{ramosaj2020cautionary}. 

Since both MICE PMM and MICE Norm rely on linear models, they cannot directly impute non-metric variables. This limitation has led to the emergence of tree-based methods that can handle mixed data types as a promising alternative. Tree-based approaches excel at handling missing data in complex, high-dimensional datasets by imputing based on observed patterns, making them effective even when missingness is not entirely random \citep{Hayes2018, HayesM2017}. They efficiently handle numerous variables and interactions without overfitting, a challenge for MICE PMM \citep{Hayes2018}. 

Three prominent tree-based imputation methods are MICE with Random Forest (RF), Chained Random Forest (\texttt{missRanger}) \citep{mayer2019missranger}, and Extreme Gradient Boosting (\texttt{mixGB}) \citep{DengL2023}. MICE with RF and Chained RF have demonstrated high imputation accuracy -- measured by the absolute distance between true and imputed data -- and strong performance in subsequent classification or prediction tasks \citep{stekhoven2012missforest, ramosaj2019predicting, thurow2021imputing, ramosaj2022relation, buczak2023analyzing}. Although their distributional accuracy for metric variables is not entirely satisfactory \citep{thurow2021imputing, thurow2024assessing}, tree-based methods have recently shown robust power properties for hypothesis testing in linear models after imputation \citep{schwerter2024evaluating}. Another (Bayesian) tree-based approach using full conditional distributions -- which in particular takes the fulfillment of nested equality and inequality edit rules for the variables into account -- is presented by \citet{assmann2024fullcond}.

This study addresses the important issue of missing data in a survey by examining how different imputation methods affect the performance of machine learning regression models in terms of accuracy and feature selection, with the latter being the main focus of the study. In official statistics, feature selection occurs when a statistical office conducts analyses itself by fitting models based on official data after imputation, or when researchers use official microdata after (suitable and legally unobjectionable) provision for scientific research via a research data center. In a simulation study, we evaluate eight advanced imputation techniques -- including three MICE variants \citep{VanBuurenGO2011}, four \texttt{missRanger} options \citep{mayer2019missranger}, and the recently proposed mixGBoost method \citep{DengL2023} -- and their compatibility with three popular algorithms \citep{GrinsztajnEA2022}: LASSO \citep{hastie2009elements}, Random Forest \citep{breiman2001random}, and XGBoost \citep{chen2016xgboost}. For comparison, we also include listwise deletion. Through a realistic simulation study, we aim to identify the optimal imputation-prediction pairings that improve feature selection and accurately reflect feature importance, thereby advancing the application of interpretable machine learning in survey analyses. This research bridges classical data science techniques with the practical challenges of missing data in empirical surveys.

Following this, the structure of this document is as follows: in Section \nameref{sec:matmet}, the dataset selected for the simulations and the techniques used to deal with missingness are presented. Then, in Section \nameref{sec:Simulation}, the design and the machine learning methods used for variable selection are described. Next, in Section \nameref{sec:modelling}, the outcomes of the simulations are detailed and discussed. Finally, the paper concludes with a summary and outlook for future work.

\section{Materials and Methods}
\label{sec:matmet}

\subsection{The NEPS Dataset}

To ensure a realistic comparison of methods, we use a simulation approach that leverages existent survey data to create simulated data, drawing inspiration from recent studies \citep{friedrich2023role, thurow2023simulate, schwerter2024evaluating}. Our data generation process for the simulations is closely aligned with the National Education Panel Study for the Starting Cohort of Adults \citep[NEPS, SC6: 13.0.0, ][]{neps_network_2022}, which examines adult education, career trajectories, and skill development in Germany for individuals born between 1944 and 1986 \citep{manual2022}. The link between income and education is of interest to the public sector for planning and evaluation purposes and is also considered in other statistics. Examples in Germany are the microcensus and the Earnings Survey,  although these set different areas of focus than the NEPS in terms of variables and type of survey.

We used data from the sixth wave of the NEPS (the 5th NEPS survey, covering data collected between 2013 and 2014), ensuring that only cases with complete data (i.e., observations with no missing entries) were included. This dataset consists of 22 variables, grouped according to the scale type for each case, as summarized in Table \ref{tab:neps_sum}.

Working hours (\textit{workinghrs}), age in years (\textit{age}), work experience in years (\textit{workexp}), and political orientation on a scale from left (0.00) to right (10.00) (\textit{leftright}) comprise four continuous variables. While these are treated as continuous, only \textit{workexp} and \textit{workinghrs} contain non-integer values in the selected dataset. The number of siblings (\textit{siblings}), number of contact attempts made before completing the survey (\textit{contactattempts}), and number of children under 6 years old in the respondent's household (\textit{children}) are three discrete variables included in the dataset.

Next, we included binary variables such as gender (\textit{gender}), fixed-term contract status (\textit{fixedterm}), participation in further training since the last interview (\textit{wb}), country of birth (Germany or not) (\textit{birthcountry}), involvement in informal learning since the last interview (\textit{ilearn}), and whether the respondent listens to classical music (\textit{classicmusic}). A ternary variable captures volunteering involvement (never, once, or at least twice) (\textit{volunteering}). Additionally, there are variables with four levels including the Classification of Profession Requirements (\textit{kldb}), marital status (\textit{maritalstatus}), highest education level attained (\textit{education}), company size (\textit{compsize}), and industry sector (\textit{sector}). There is also a five-level variable that captures parental school qualification (\textit{schoolparents}) and the federal state of residence at the time of the survey (\textit{fedstate}). Detailed information for factor variables with four or more levels is provided in Appendix \ref{appendix:factors}. All the factor variables were encoded following a \textrm{one-hot encoding} approach (also known as \textrm{dummy encoding}): it replaces a $k$-level variable with $k-1$ binary indicator features by removing the most frequent category in the variable. This is in order to avoid collinearity with an intercept \citep[p ~126]{james2023introduction}. Table \ref{tab:neps_sum} shows in bold all the levels that were removed for each factor variable after the \textrm{one-hot encoding} was applied to the dataset. 

Finally, the target variable is \textit{ln\_real\_inc}, representing the natural logarithm of the respondent's monthly income, serving as the continuous outcome variable. For a comprehensive understanding of the variables and their distributions, histograms or count plots for each variable are available in Appendix \ref{appendix:distplots}.

In total, the dataset used for our evaluation has 3,886 complete observations of 49 features, of which 4 are metric, 3 are integer and 42 binary variables stemming from the factor encoding. The 50th variable is the outcome.

\begin{table}[htb!]
		\caption{Variables from the NEPS Dataset with their Scale Type and their most Important Summary Statistics}
			\begin{tabulary}{\textwidth}{lslL}
				\toprule
				\textbf{Variable} & \textbf{Type} & \textbf{Type} & \textbf{Statistics} \\
				\midrule
                \textbf{Metric variables}\\ 
				\cmidrule{1-1}
				\textit{workinghrs} & numeric & Continuous & Min: 0.00, Median: 40.00, Max: 90.00 \\
				\textit{age} & numeric & Continuous & Min: 26.00, Median: 49.00, Max: 69.00 \\ 
				\textit{workexp} & numeric & Continuous & Min: 0.50, Median: 25.00, Max: 55.58 \\ 
				\textit{leftright} & numeric & Continuous & Min: 0.00, Median: 5.00, Max: 10.00 \\ 
				
				& & & \\ [-7pt]
				\textbf{Integer variables}  \\ 
				\cmidrule{1-1}
				\textit{siblings} & numeric & Discrete & Min: 0.00, Median: 1.00, Max: 23.00 \\ 
				\textit{contactattempts} & integer & Discrete & Min: 1, Median: 6, Max: 117 \\ 
				\textit{children} & integer & Discrete & Min: 0, Median: 0, Max: 3 \\ 
				
				& & & \\ [-7pt]
				\textbf{Factor variables}  \\ 
				\cmidrule{1-1}
				\textit{gender} & factor & Factor - 2 levels & \textbf{male: 1971}, female: 1915 \\ 
				\textit{fixedterm} & factor & Factor - 2 levels & \textbf{no: 3643}, yes: 243 \\ 
				\textit{wb} & factor & Factor - 2 levels & \textbf{no: 2802}, yes: 1084 \\ 
				\textit{birthcountry} & factor & Factor - 2 levels & \textbf{in Germany: 3647}, abroad: 239 \\ 
				\textit{ilearn} & factor & Factor - 2 levels & \textbf{yes: 2743}, no: 1143 \\ 
				\textit{classicmusic} & factor & Factor - 2 levels & \textbf{yes: 2094}, no: 1792 \\ 
				\textit{volunteering} & factor & Factor - 3 levels & \textbf{no: 1875}, 2orMore: 1482, once: 529 \\ 
				\textit{kldb} & factor & Factor - 4 levels & \textbf{Skilled: 1861}, HighComplx: 1162, Complx: 639, LowComplx: 224 \\ 
				\textit{maritalstatus} & factor & Factor - 4 levels & \textbf{married: 2742}, single: 804, divorced: 286, widowed: 54 \\
				\textit{education} & factor & Factor - 4 levels & \textbf{Secondary: 1336}, University: 1293, Abitur: 677, LowerSec: 580 \\ 
				\textit{compsize} & factor & Factor - 4 levels & \textbf{3: 1465}, 4: 1431, 1: 589, 2: 401 \\ 
				\textit{sector} & factor & Factor - 4 levels & \textbf{3: 1506}, 2: 1187, 1: 1151, 4: 42 \\ 
				\textit{schoolparents} & factor & Factor - 5 levels & \textbf{HauptS: 2120}, HighSDip: 936, RealS: 780, NoSCert: 43, Other: 7 \\ 
				\textit{fedstate} & factor & Factor - 16 levels & \textbf{NW: 845}, BY: 622, BW: 457, NI: 421, HE: 332, RP: 215, SN: 205, BE: 136, ST: 124, TH: 114, BB: 113, SH: 109, HH: 72, MV : 55, SL: 46, HB: 20 \\ 
				
				& & & \\ [-7pt]
				\textbf{Target variable}  \\ \cmidrule{1-1}
				\textit{ln\_real\_inc} & numeric & Continuous & Min: 3.83, Median: 8.04, Max: 9.97 \\ 		
				\bottomrule
			\end{tabulary}
   \caption*{\footnotesize \textit{Note:} For numeric and integer variables: minimum, median and maximum values are shown. For factors: all the levels with their count in decreasing order and the most frequent in \textbf{bold} letters.}  \label{tab:neps_sum}
\end{table}

\subsection{Methods to Handle Missing Data}

The investigated imputation methods cover listwise deletion, three MICE options \citep{VanBuurenGO2011}, four \texttt{missRanger} options \citep{stekhoven2012missforest, mayer2019missranger} as well as the recently proposed mixGBoost imputation approach \citep{DengL2023}. 

\textbf{Listwise Deletion}, or complete-case analysis, removes all observations with any missing values in either the predictors or the outcome. This method is simple to implement, requiring no data modification or imputation, and produces unbiased estimates if data are missing completely at random (MCAR). However, it can lead to significant information loss, especially with many variables affected by non-response, and may result in biased estimates if the data are not MCAR \citep{vanBuuren2018,VanGinkelEA2020, schwerter2024evaluating}. 

In contrast to this rather simple benchmark approach, we considered state-of-the-art prediction-model-based imputation approaches. These treat the imputation of missing values as a prediction problem, where the variable with missing data is predicted using the remaining variables as predictors.
To this end, several statistical and ML-based prediction models are in use. In the present paper, we consider approaches based on (Bayesian) linear models, Random Forests and XGBoost. We start by explaining the former, which usually appear in the context of Multiple Imputations.

\textbf{Multiple Imputation (MI)} addresses the problem of missing data by generating several plausible values for each missing entry rather than a single estimate. By repeatedly imputing missing values based on all available information and creating multiple imputed versions of the dataset, MI allows for subsequent analysis with measurable uncertainty, accounting for variability in the imputation \citep{rubin1996multiple, raghunathan2015missing}. Currently, \textbf{Multivariate Imputation by Chained Equations (MICE)} \citep{VanBuurenGO2011} is one of the most popular MI algorithms that can impute multiple variables simultaneously. We study three different MICE approaches:

\textbf{MICE Norm} is a 
Bayesian imputation approach under the normal linear model which is based on early ideas from \cite{rubin1987multiple}. Thereby, the observed data is used to estimate the posterior distribution from which the parameters of the linear model are drawn, which in turn is used for the prediction-model-based imputation. 

\textbf{MICE PMM} extends MICE Norm by selecting a set of so-called candidate donors (by default: $k=5$) from the observed data whose outcome values are closest to the predicted ones (obtained from the Bayesian linear model). One donor is randomly chosen, ensuring that only observed outcome values are imputed.

The MICE package \citep{VanBuurenGO2011} also contains the \textbf{MICE RF} algorithm \citep{doove2014recursive}. It uses $k$ individual decision trees trained on bootstrap samples from the observed data. For each missing value, the data point is passed through all trees, landing in a terminal node for each. A random donor is selected from the observations in these nodes, and one donor is randomly chosen from the $k$ donors for imputation.

Another popular and highly cited (almost $5,000$ Google Scholar citations at the time of submission) imputation method is given by the missForest algorithm proposed in \cite{stekhoven2012missforest}, which can also be used with mixed-type data. 
It iteratively refines initial imputations using Random Forests as prediction models. As the process progresses, previously imputed values are used for further (improved) predictions. In the present paper, we use the computationally fast \textbf{missRanger} implementation \citep{mayer2019missranger}.

Unlike MICE RF, the original missForest algorithm doesn't rely on reflecting uncertainty in its imputation mechanism. To counteract this somewhat, the implementation \cite{mayer2019missranger} additionally offers \textbf{missRanger with predictive mean matching} as an option. 
This not only ensures that imputation is restricted to observed values but also tries to better reflect the underlying (but unknown) variable distributions with the imputed values. Since there are no default values for the number of donors $k$ when applying \texttt{missRanger}, we explore three typical options: $k=3,5,10$. In addition, the choice $k=0$ gives the missForest implementation. 

Similar to Random Forest-based imputation approaches, \cite{DengL2023} introduced \texttt{mixgb}, an XGBoost-based imputation method \citep{chen2016xgboost}. It can also handle mixed-type data and offers predictive mean matching. We use the default values for predictive mean matching as recommended in \citep{DengL2023}: $k=5$ for the imputation of continuous variables and no PMM in the case of categorical data.

\section{Simulation}
\label{sec:Simulation}

The study employs a nested resampling approach initiated with an outer 3-fold cross-validation (see Figure \ref{fig:design} for a graphical overview). For each fold, the dataset is partitioned into training ($\mathcal{D}^{train}_{out\mathbf{f}}$) and test sets ($\mathcal{D}^{test}_{out\mathbf{f}}$). The following sequence is applied to each outer fold with 100 iterations:

\textbf{Amputation and Imputation}: Missing data is introduced following an MAR mechanism as described by \cite{thurow2021imputing}, with missing rates of 10\%, 30\%, 50\%, and 70\%. Subsequently, imputation methods listed in Table \ref{tab:impMethods} are applied with default \texttt{R} package parameters with $m = 5$ imputations. After imputation, metrics such as NRMSE, IPM, and correlation distances are recorded. Given the focus on feature selection, these results together with the definition of the metrics are only reported in the Appendix \ref{appendix:addResults}.

\begin{table}[htb!]
		\caption{\label{tab:impMethods}The Nine Imputation Methods under Investigation with the Chosen Number of Multiple Imputations \texttt{m} and the used \texttt{R} Functions and Parameters for Their Implementation} 
		\centering
		\small
		\begin{tabulary}{\textwidth}{lcL}
			\toprule
			\textbf{Method} & \textbf{$m$} & \textbf{R Function and Parameters} \\
			\midrule
			Listwise (\texttt{listwise}) & 0 &  \texttt{stats::complete.cases()} \\
			Bayesian Linear (\texttt{norm}) & 5 & \texttt{mice(method = "norm")}\\
			Predictive Mean Matching (\textit{pmm}) & 5 & \texttt{mice(method = "pmm")}\\
			Random Forests (\texttt{RF})& 5 & \texttt{mice(method = "rf")}\\
			Chained RF (\texttt{missRanger0}) & 5 & \texttt{missRanger(pmm.k = 0)} + replicate(5,...)\\
			Chained RF (\texttt{missRanger3}) & 5 & \texttt{missRanger(pmm.k = 3)} + replicate(5,...)\\
			Chained RF (\texttt{missRanger5}) & 5 & \texttt{missRanger(pmm.k = 5)} + replicate(5,...)\\
			Chained RF (\texttt{missRanger10}) & 5 & \texttt{missRanger(pmm.k = 10)} + replicate(5,...)\\
			XGBoost (\texttt{mixGB}) & 5 & \texttt{mixGB()}\\
			\bottomrule
		\end{tabulary}
	\end{table}

\textbf{Hyperparameter Tuning and Model Evaluation}: An inner 4-fold cross-validation is employed for hyperparameter tuning for each statistical learner, optimizing for MSE. The optimal hyperparameters are then used to train the five multiply imputed training datasets. The best model is applied on the respective test set, yielding $m = 5$ \textit{MSE} values. To analyze the effect on feature selection, we save the best feature sets per multiply imputed training data. The details on the process followed for feature selection is presented with a more detailed explanation below. Moreover, a single \textit{MSE} is derived by averaging the pooled MSEs across folds and, again, given the focus on feature selection, these results are only reported in the Appendix \ref{appendix:addResults}. 

\textbf{Performance}: To establish a performance baseline and explore regression capabilities on the full dataset, the entire procedure is also conducted without the amputation, providing benchmark results with the complete data. All tasks are executed using the \texttt{mlr3} framework \citep{Bischl2024} in \texttt{R} \citep{R2022}.

\subsection{Hyperparameter Tuning and Feature Selection}
\label{ssec:tune}

To efficiently perform hyperparameter tuning for LASSO, Random Forest and XGBoost, a nested cross-validation with four inner folds is applied to each imputed dataset (see the red loops for the Tuning step as depicted in Figure \ref{fig:design} for a graphical overview). This evaluation is conducted using grid search with a resolution and batch size of 5, implemented via the method \texttt{mlr3tuning:tnr} within the automatic tuning process \texttt{mlr3tuning:auto\_tuner}, using Mean Squared Error (MSE) as the loss function to identify the optimal set of parameters.

For the Lasso Regularized Linear Model, we utilize the \texttt{regr.glmnet} learner from \texttt{mlr3learners} \citep{lasso}. Parameters include \texttt{alpha = 1} to enable only the LASSO penalty, \texttt{nfolds = 5} for a 5-fold cross-validation method to tune the parameter $\lambda$, and \texttt{lambda.min.ratio} set to be tuned in the interval between 0.005 and 0.03, where MSE results converge with the complete dataset. If a variable coefficient is not set equal to zero, the variable is selected by LASSO.

For Random Forest, the \texttt{regr.ranger} model from \texttt{mlr3learners} uses the \texttt{ranger} package, implementing the method described by \cite{ranger}. As a compromise between accuracy and run-time, we used 200 trees for the implementation of Random Forest. Based on the hyperparameter tuning suggestions by \cite{probst2019hyperparameters} and the best results obtained for the \textit{complete} dataset, the following hyperparameters are used: \texttt{mtry.ratio = \{0.15, 0.3, 0.45\}} to vary the number of columns used to train each tree between 
$\sqrt{p}$ and $p/3$ for $p=50$ (= 49 predictors + 1 imputation indicator); \texttt{sample.fraction = [0.2, 0.9]} to vary the fraction of observations used to train the trees between 20\% and 90\%; and \texttt{min.node.size = \{5, 7, 10\}} to set the minimum number of observations per leaf. Unlike linear regression models, Random Forests do not deliver concrete parameter estimates but only feature importances of variables (\texttt{importance = "impurity"}) that can be used for feature / model selection. If the variable importance was higher than the median importance, we counted the feature as selected, following the filter-based Feature Selection approach introduced by \cite[][Section 6.1.4]{Bischl2024}.

For XGBoost models, \texttt{regr.xgboost} from \texttt{mlr3learners} is employed. Following the tuning suggestions by \cite{putatunda2018comparative} and \cite{srinivas2022hyoptxg}, as well as testing on the \textit{complete} dataset, the tuned parameters include: \texttt{eta = [0.1, 1]} chosen to form a grid of values on a logarithmic scale, controlling the learning rate or shrinkage parameter; \texttt{nrounds = [50, 100]} controlling the number of trees; \texttt{maxdepth = \{1, 5, 10\}} controlling tree depth; and \texttt{colsample\_bytree = \{0.2, 0.5, 0.75, 1\}} controlling the percentage of parameters $p$ used as candidates to train the trees. As with Random Forests, if the variable importance was higher than the median importance, we counted the feature as selected. Therefore, the hyperparameter tuning step involves evaluating five combinations for LASSO, 45 for Random Forests, and 120 for XGBoost. 

After tuning and evaluation, we end up with $m = 5$ models, each with possibly distinct features that, following the pooling approach presented by \cite{zahid2020variable}, akin to pooling metrics for feature selection after multiple imputation, is used to choose the most important among them as follows:
\begin{equation}
	\text{select} \, X_j, \quad \text{if} \quad \frac{\sum_{t=1}^m \mathbb{1}(X_{j,t} \text{ is chosen})}{m} \geq \delta, \label{eq1}
\end{equation}
where $\mathbb{1}(X_{j,t} \text{ is chosen})$ is one if feature $X_j$ is chosen for iteration (or imputation) $t$, and zero otherwise.  A feature is selected if it appears in a proportion greater than or equal to $\delta = 0.6$ \citep{zahid2020variable}.
Thus, for $m = 5$, a feature is chosen if retained in at least three of the five models. 

Afterward, resulting in one model per outer fold, the selection undergoes another pooling between the 3 feature subsets, applying the same $\delta = 0.6$ to select features present in at least two out of the three folds. 
We then sum up how often per simulation run a variable was selected to compare the results across the 100 simulations.

\usetikzlibrary {arrows.meta}

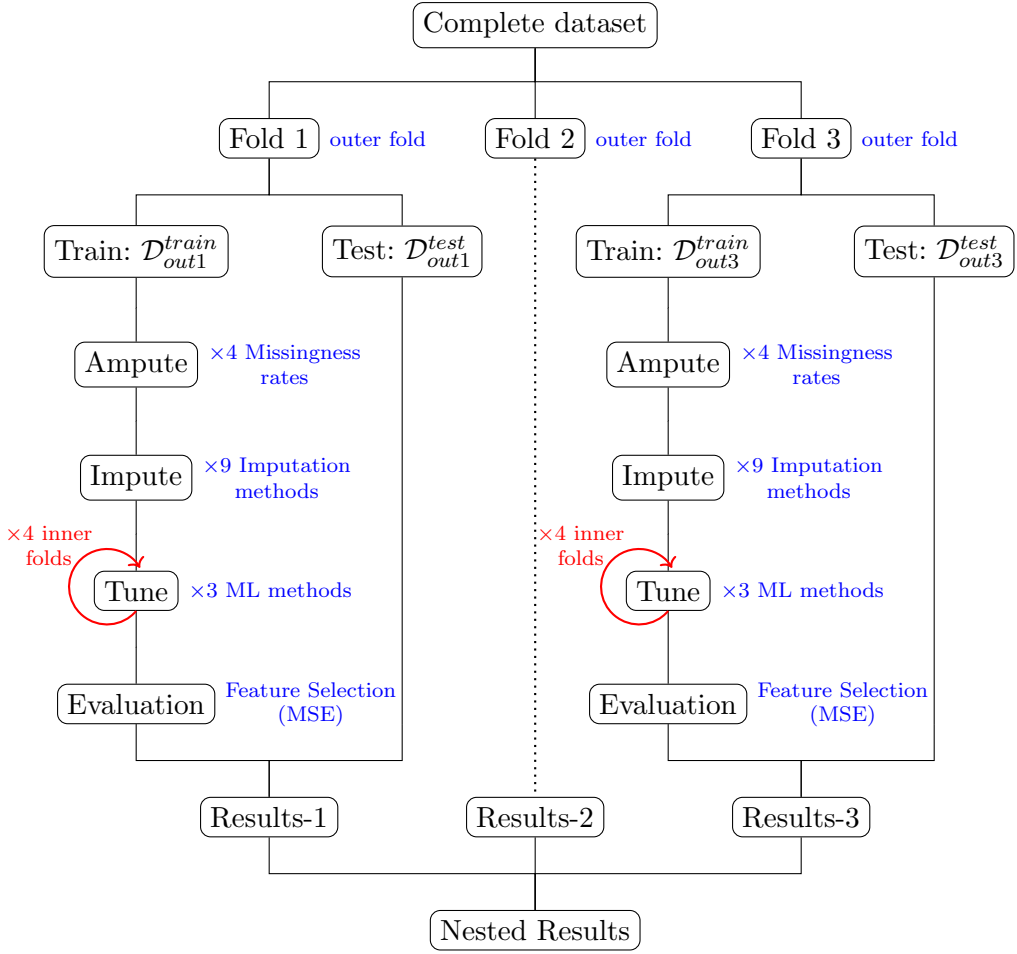
\begin{figure}[H]
	\caption{Overview of our simulation Design with Nested Cross Validation}
	\label{fig:design}
	\begin{center}
		\begin{tikzpicture} [every node/.style={rectangle,draw,rounded corners=.8ex}, level distance=15mm,
			font = \normalsize]
			\node (top) at (0,12.0) {Complete dataset} [grow = down, edge from parent fork down, sibling distance = 3.5cm]
			child {node [label= {[blue] right:\scriptsize outer fold }] {Fold 1 }
				child {node {Train: $\mathcal{D}^{train}_{out1}$}
					child {node [label= {[blue] right:{\scriptsize \shortstack{$\times$4 Missingness\\ rates }}}] {Ampute}
						child {node [align=left, label={[blue] right:{\scriptsize \shortstack{$\times$9 Imputation\\methods}}}] {Impute}
							child {node [align = left, label= {[blue] right:{\scriptsize \shortstack{$\times$3 ML methods}}}] (tune1) {Tune}
								child{node [label= {[blue] right:{\scriptsize \shortstack{Feature Selection \\ (MSE)}}}] (eval1) {Evaluation}}
							}
						}
					} 
				}
				child {node[xshift=0pt] (test) {Test: $\mathcal{D}^{test}_{out1}$}}
			}
			child {node (f2) [label= {[blue] right:\scriptsize outer fold }] {Fold 2}}
			child {node (f3) [label= {[blue] right:\scriptsize outer fold }] {Fold 3}
				child {node {Train: $\mathcal{D}^{train}_{out3}$}
					child {node [label= {[blue] right:{\scriptsize \shortstack{$\times$4 Missingness\\ rates }}}] {Ampute}
						child {node [label= {[blue] right:{\scriptsize \shortstack{$\times$9 Imputation\\methods}}}] {Impute}
							child {node [label= {[blue] right:{\scriptsize \shortstack{$\times$3 ML methods}}}] (tune3) {Tune}
								child{node [label= {[blue] right:{\scriptsize \shortstack{Feature Selection \\ (MSE)}}}] (eval3) {Evaluation}}
							}
						}
					} 
				}
				child {node[xshift=0pt] (test2) {Test: $\mathcal{D}^{test}_{out3}$}}
			}
			;
			
			\node (bottom) at (0,0) {Nested Results} [grow = up, edge from parent fork up, sibling distance = 3.5cm]
			child{node (mod3) {Results-3}
				child{node[transparent] (test_f2){Test}}
				child{node[transparent]  {Train}}
			}
			child{node (mod2) {Results-2}}
			child{node (mod1) {Results-1}
				child{node[transparent] (test_f){Test}}
				child{node[transparent]  {Train}}
			}
			;
			
			\draw (test) -- (test_f.south);
			\draw (test2) -- (test_f2.south);
			\draw [thick, dotted] (f2) -- (mod2);
            \draw [red, ->, line width = 0.8] (tune1.south)arc(320:30:0.5);
            \draw [red, ->, line width = 0.8] (tune3.south)arc(320:30:0.5);

            \node[draw=none, red] at (-6.4,5.1) {\scriptsize \shortstack{$\times$4 inner\\folds}};
            \node[draw=none, red] at (0.6,5.1) {\scriptsize \shortstack{$\times$4 inner\\folds}};
		\end{tikzpicture}
	\end{center}
\end{figure}

\section{Feature Selection Results}
\label{sec:modelling}

Here, we present the results for each learner when paired with one of the 9 imputation methods, beginning with the feature selection for LASSO (Figures~\ref{fig:featsLMa}-\ref{fig:featsLMb}), followed by the feature importances of Random Forests (Figures~\ref{fig:featsRFa}-\ref{fig:featsRFb}), and XGBoost (Figures~\ref{fig:featsXGBa}-\ref{fig:featsXGBb}).
For each learner, we first show the results for true positives (Figures~\ref{fig:featsLMa}, \ref{fig:featsRFa} and \ref{fig:featsXGBa}) with increasing missing rates (from left $10\%$ to right $70\%$). Thereby, a deeper shade of blue signifies more frequent selection or importance of a feature across the 100 iterations. For false positives (Figures~\ref{fig:featsLMb}, \ref{fig:featsRFb}, and \ref{fig:featsXGBb}), a red color scheme is employed to this end.

The focus of this study is on feature selection. However, additional results regarding imputation accuracy (NRMSE, IPM, and correlation) and predictive performance of the regression learners (MSE) are shown in the Appendix \ref{appendix:addResults} and are only briefly addressed in this article.

\subsection{LASSO}
 
\begin{figure}[htb!]
	\centering
	\caption{Feature Selection for the Lasso Penalized Linear Models: True Positives}
	\label{fig:featsLMa}
	\includegraphics[width = \textwidth]{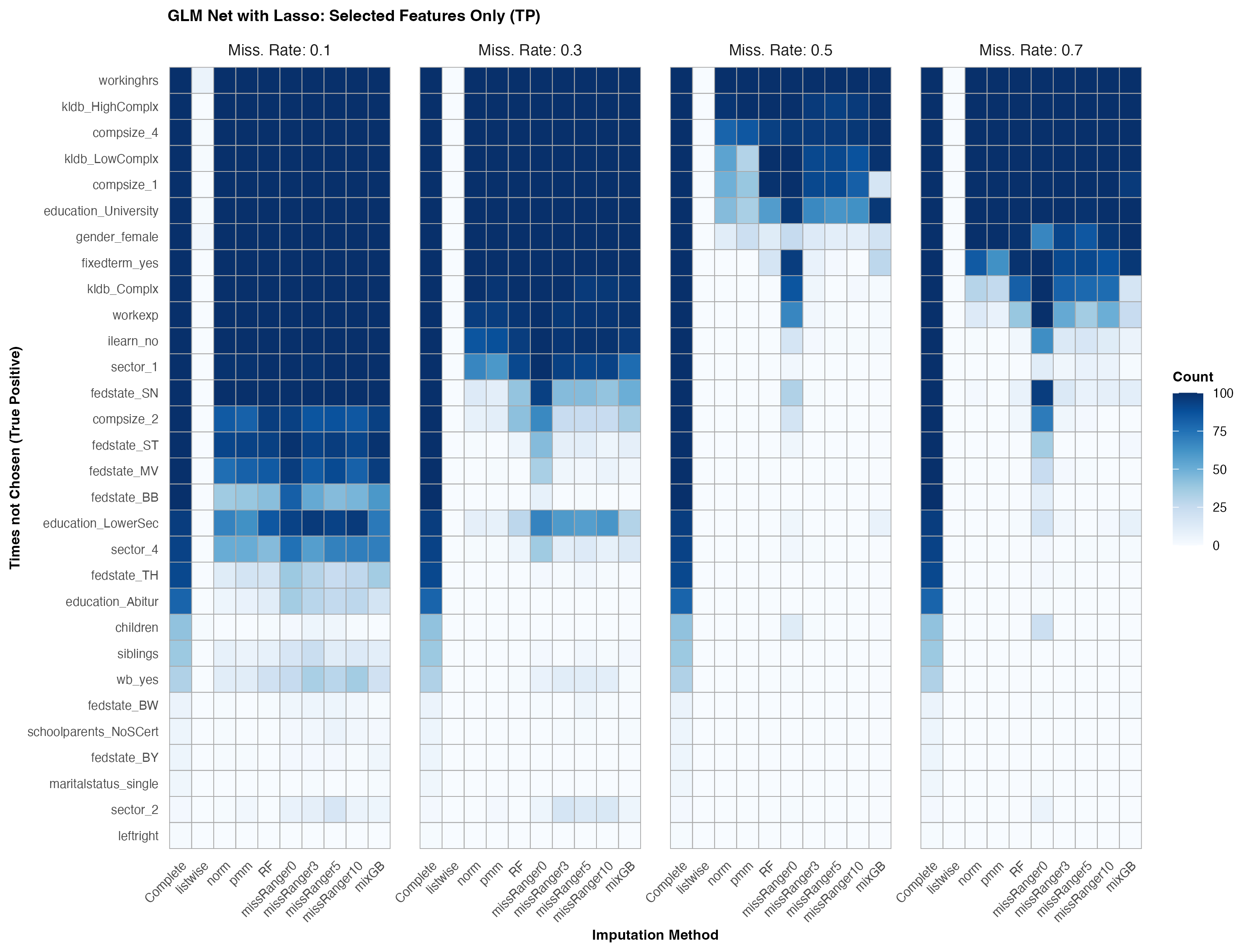}
\end{figure}

\textbf{True Positives:} Figure \ref{fig:featsLMa} shows the frequency with which each feature was selected by LASSO during the prediction of respondents' income across 100 iterations for both the complete dataset and the various imputed datasets. For the complete data (column to the left as a comparison in each subpanel),
30 features (corresponding to the number of displayed rows) were selected from the 50 (49 original plus 1 imputation indicator) features in at least one of the 100 iterations by LASSO. 

When comparing the imputation methods (represented by the other nine columns in each subpanel), \texttt{listwise deletion} is often unable to perform feature selection, as explained in the previous section. It is therefore excluded from the following discussion, which focuses solely on the remaining eight imputation approaches. Among these, \texttt{missRanger0} stands out as the most effective at selecting features compared to the baseline (Complete), particularly at the two highest missing rates. 

For the 10\% missingness case, nearly half of the features selected in the baseline are retained by almost all imputation methods with high selection frequencies (more than 50 times across 100 runs). Among the 20 features that were almost always selected by the baseline, only 13 were also (almost) always selected after the amputation and imputation process. For the next eight most frequently selected features by the baseline, \texttt{missRanger} without PMM, and then MICE RF, \texttt{missRanger} with PMM and \texttt{mixGB} 
demonstrated greater alignment with the feature selection patterns observed in the complete data. MICE Norm and MICE PMM showed less alignment (see particularly the features\textit{compsize\_2, fedstate\_ST, fedstate\_MV, fedstate\_BB, education\_LowerSec,  sector\_4}).

In cases where the LASSO selection frequency for the complete data was significantly below 100 (e.g., for features from \textit{children} to \textit{leftright}), the selection frequency after the imputation consistently remained below 100 as well. This indicates that the amputation and imputation process did not lead to any noticeable improvement in the selection.

This pattern holds true across the other three missing rates with only minor exceptions, generally showing a clear reduction in the feature selection frequency. At the 30\% missing rate, ten features were still selected in each iteration after imputation. However, this number drops to three at the 50\% missing rate and six at the 70\% missing rate. That is, the increase in missingness leads to a reduction in the number of chosen features until the 50\% rate and then slightly increases again at 70\%. However, the feature selection frequency is much lower than observed at the 30\% case.

Irrespective of the missing rate, most true positive features are selected when missing values were imputed using \texttt{missRanger} without PMM. Overall, MICE Norm appears to perform the worst among the eight imputation methods in enabling LASSO to retain the same features as in the complete dataset. 

Only three features: \textit{workinghrs}, \textit{kldb\_HighComplx} and \textit{compsize\_4} were chosen very frequently by all models and at any missing rate.

\begin{figure}[htb!]
	\centering
	\caption{Feature Selection for the Lasso Penalized Linear Models: False Positives}
	\label{fig:featsLMb}
	\includegraphics[width = \textwidth]{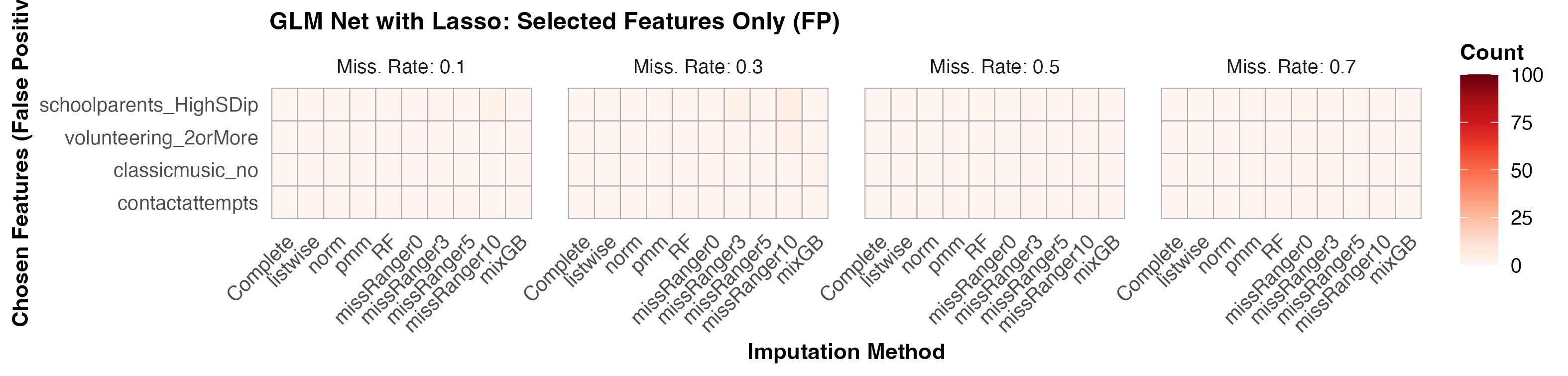}
\end{figure}

\textbf{False Positives:} The number of false positives, defined as selected features after the amputation and imputation process that were not selected for the complete data (baseline), is nearly zero for LASSO (see Figure \ref{fig:featsLMb}). Only the different \texttt{missRanger} variants and \texttt{mixGB} selected a false positive in at most four out of 100 iterations for the 10\%, 30\% and 70\% missing rate: the feature \textit{schoolparents\_HighSDip} was only chosen in four iterations after imputing with \texttt{missRanger10}. Thus, compared to the baseline, LASSO seems to select a sparse set of variables after amputation and imputation with hardly any false positives.

\subsection{Random Forests} 

\begin{figure}[htb!]
	\centering
	\caption{Feature Selection for the Random Forests: True Positives}
	\label{fig:featsRFa}
	\includegraphics[width = \textwidth]{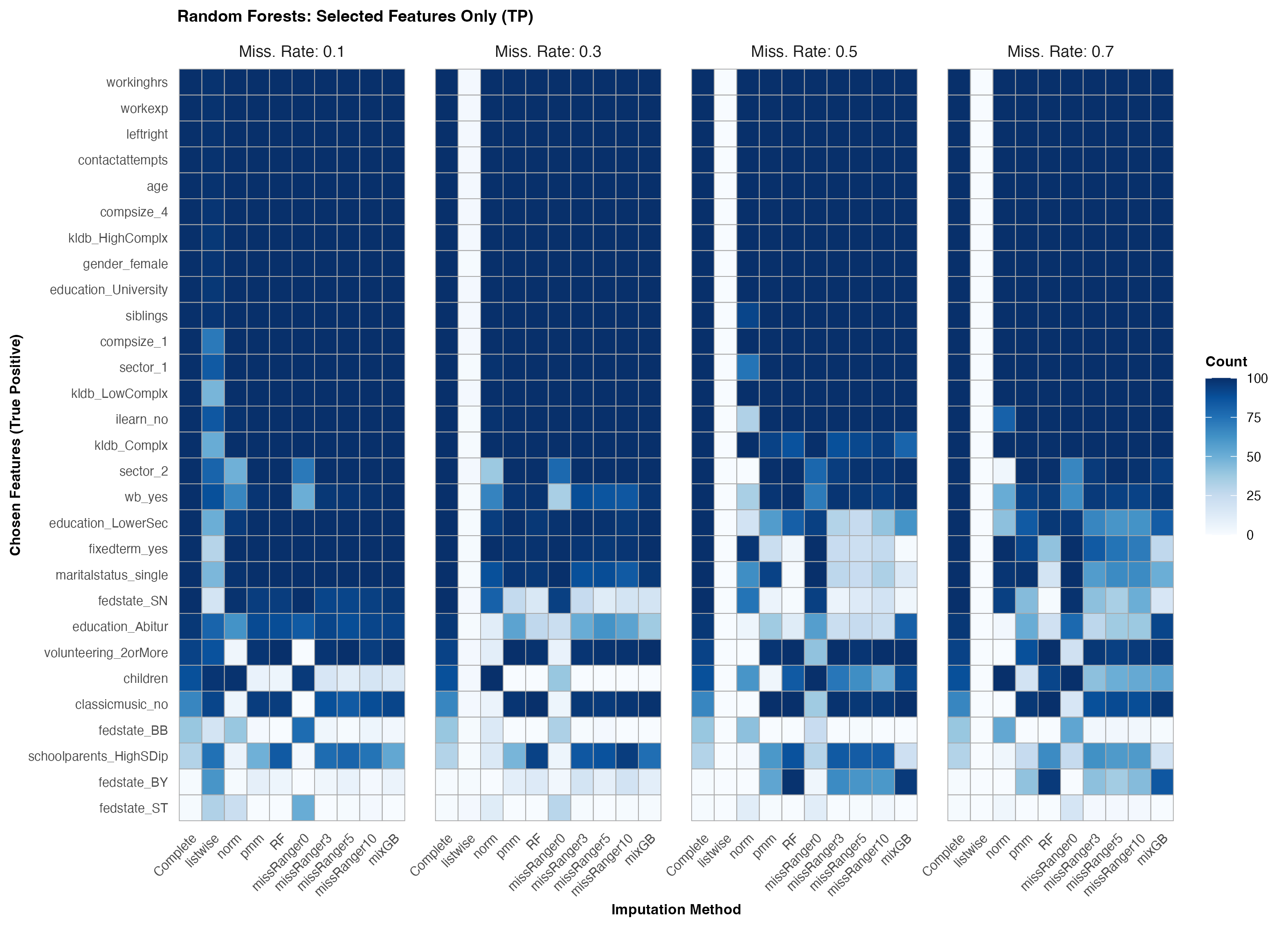}

\end{figure}

\textbf{True Positives:} Figure \ref{fig:featsRFa} highlights two important differences between Random Forests and LASSO with respect to feature selection. The sensitivity of feature selection to the missing rate is less pronounced in Random Forests than in LASSO. In particular, almost half of the features selected in the baseline cases are consistently selected across all imputation models, with a frequency greater than 50 out of 100 iterations, regardless of the proportion of missing data in the dataset.

Interestingly, while 29 features were selected at least once using the Random Forest on the complete dataset, five features not selected by the LASSO (Figure~ \ref{fig:featsLMa}) were included in some of the Random Forests: \textit{contactattempts, volunteering\_2orMore, classicmusic\_no, schoolparents\_HighSDip} and \textit{age}. Among these, the first four were identified as false positives in some LASSO selections (Figure~\ref{fig:featsLMb}), whereas \textit{age} was not chosen by LASSO in any case. The \textit{imputed} feature was consistently excluded in both baseline model sets.

When examining the effect of the missing rate on the selection of features for Random Forests, the results are similar to those of LASSO, although the variation in the number of features at different rates is less pronounced. The number of features selected is lowest at 50\%, while the results at the 30\% and 70\% missing rates are more similar. When comparing imputation models, the tendency for one method to select significantly more or fewer true positive features is less pronounced with Random Forests than with LASSO. 

In general, the methods show consistent and similar behavior at the same missing rate. Interestingly, even the \texttt{listwise deletion} approach achieves considerable feature selection at the 10\% missing rate. However, 
some of the results differ from the baseline case. Moreover, for higher missing rates, \texttt{listwise deletion} did not lead to feature selection in almost all cases (with maximum 2 times out of 100 for 30\% and none for the 2 highest rates). At the two highest missing rates, the three \texttt{missRanger} models with PMM struggled to select features with the same frequency in the bottom half of commonly selected features compared to baseline.

\begin{figure}[htb!]
	\centering
	\caption{Feature Selection for the Random Forests: False Positives}
	\label{fig:featsRFb}
	\includegraphics[width = \textwidth]{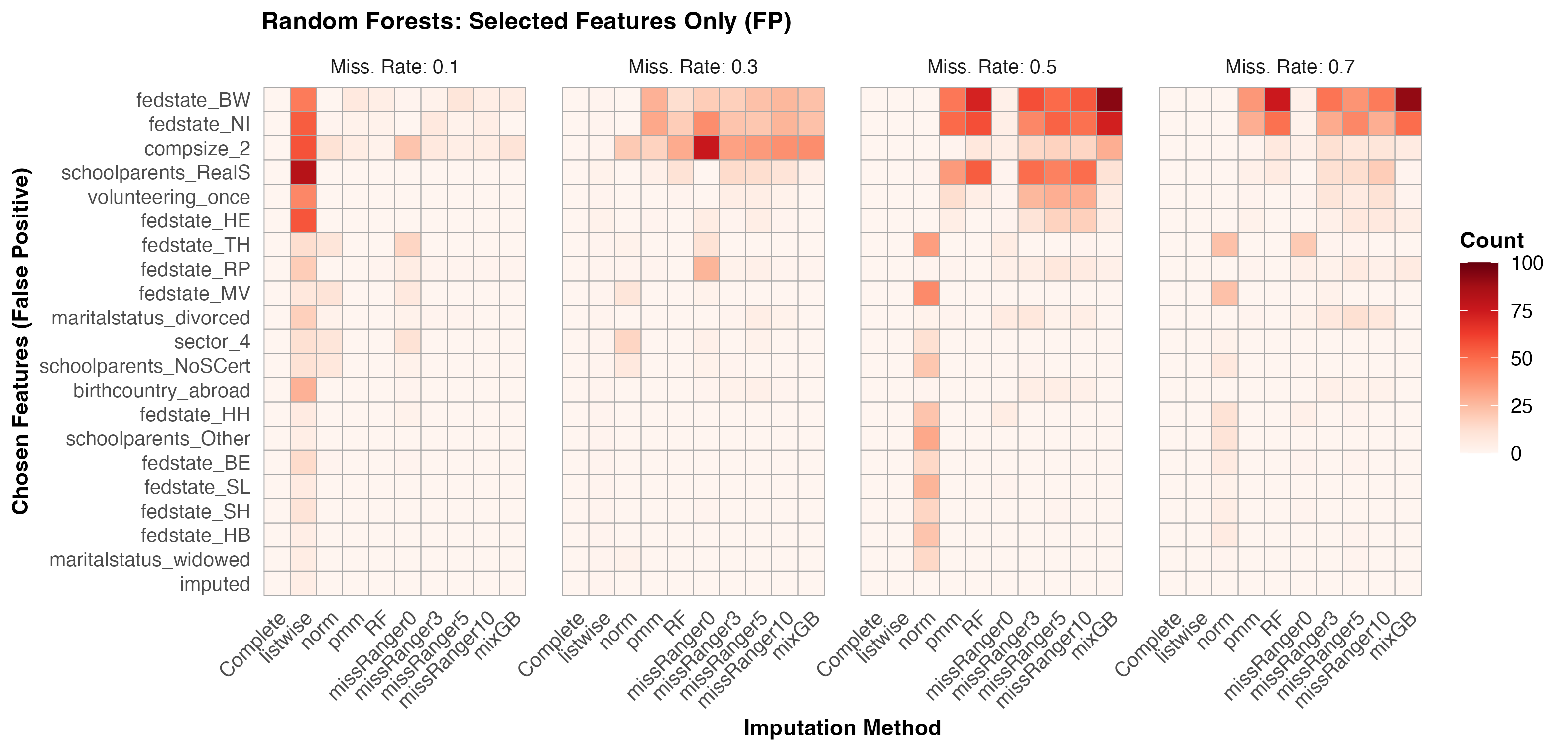}
\end{figure}

\textbf{False Positives:} Investigating the false positives shown in Figure \ref{fig:featsRFb}, it is clear that all of the remaining features are selected by at least one model, with significantly higher selection frequencies compared to LASSO. The most striking example is the feature \textit{fedstate\_BW}, which is incorrectly selected by Random Forests more than 90 times for the two highest missing rates when imputation is done with \texttt{mixGB}.

The influence of the missing rate on the erroneous selection of features using Random Forests manifests itself as an increase in false positives up to a missing rate of 50\%, followed by a decrease at 70\%. An exception is \texttt{listwise deletion}: Here, the Random Forest consistently selects more false positives at the lowest missing rate of 10\%. With increasing missing rates up to 50\%, Random Forest is particularly prominent in selecting several features that were not considered relevant in the baseline cases for \texttt{MICE Norm}. For the tree-based methods MICE RF, \texttt{mixGB} and the \texttt{missRanger} approaches with PMM, only three to five features were wrongly selected with a high frequency, at missing rates from 30\% to 70\%. For \texttt{missRanger} without PMM, this was only the case at the 30\% missing rate.

\subsection{XGBoost}

\begin{figure}[htb!]
	\centering
	\caption{Feature Selection for XGBoost: True Positives}
	\label{fig:featsXGBa}
	\includegraphics[width = \textwidth]{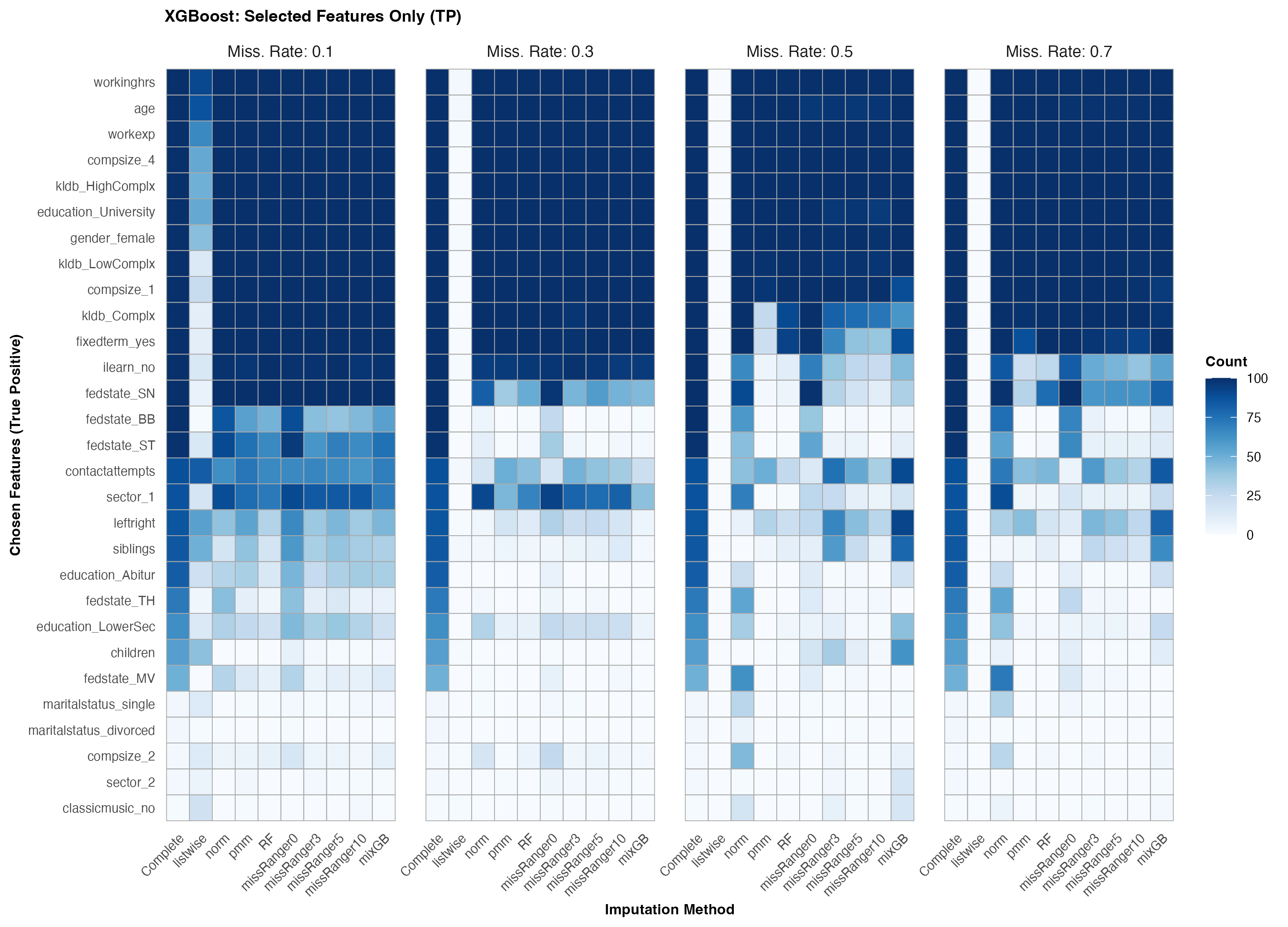}
  \caption*{{\footnotesize{\textit{Note}:}}}

\end{figure}

\textbf{True Positives:} At first glance, Figure \ref{fig:featsXGBa} shows that XGBoost exhibits greater selectivity in feature selection compared to the Random Forest, reflecting its improved ability to filter and select features \citep{chen2016xgboost}. Although 29 features were selected in at least one of the 100 baseline simulations (i.e., using XGBoost on the complete data), the specific features selected differ and are selected less frequently compared to Random Forests. In particular, nine features were consistently selected by XGBoost across all evaluated missing rates using all imputation models.

Regarding the impact of missingness on feature selection, the results were not significantly different from those observed with the other two learners. As the missingness increases, fewer features are selected up to a missing rate of 50\%. At 70\%, the number of selected features slightly increases, but does not reach the level observed at 30\%.

Comparing the different imputation models, \texttt{listwise deletion} only has a non-negligible feature selection at the 10\% missing rate and again mostly fails to capture the same features as those identified in the baseline runs. In addition, \texttt{MICE Norm} emerges as the method that selects more features overall, with some features being selected even more frequently than in the baseline. This is in contrast to the results for the LASSO and (but to a lesser extent) the Random Forest.

\begin{figure}[htb!]
	\centering
	\caption{Feature Selection for XGBoost: False Positives}
	\label{fig:featsXGBb}
	\includegraphics[width = \textwidth]{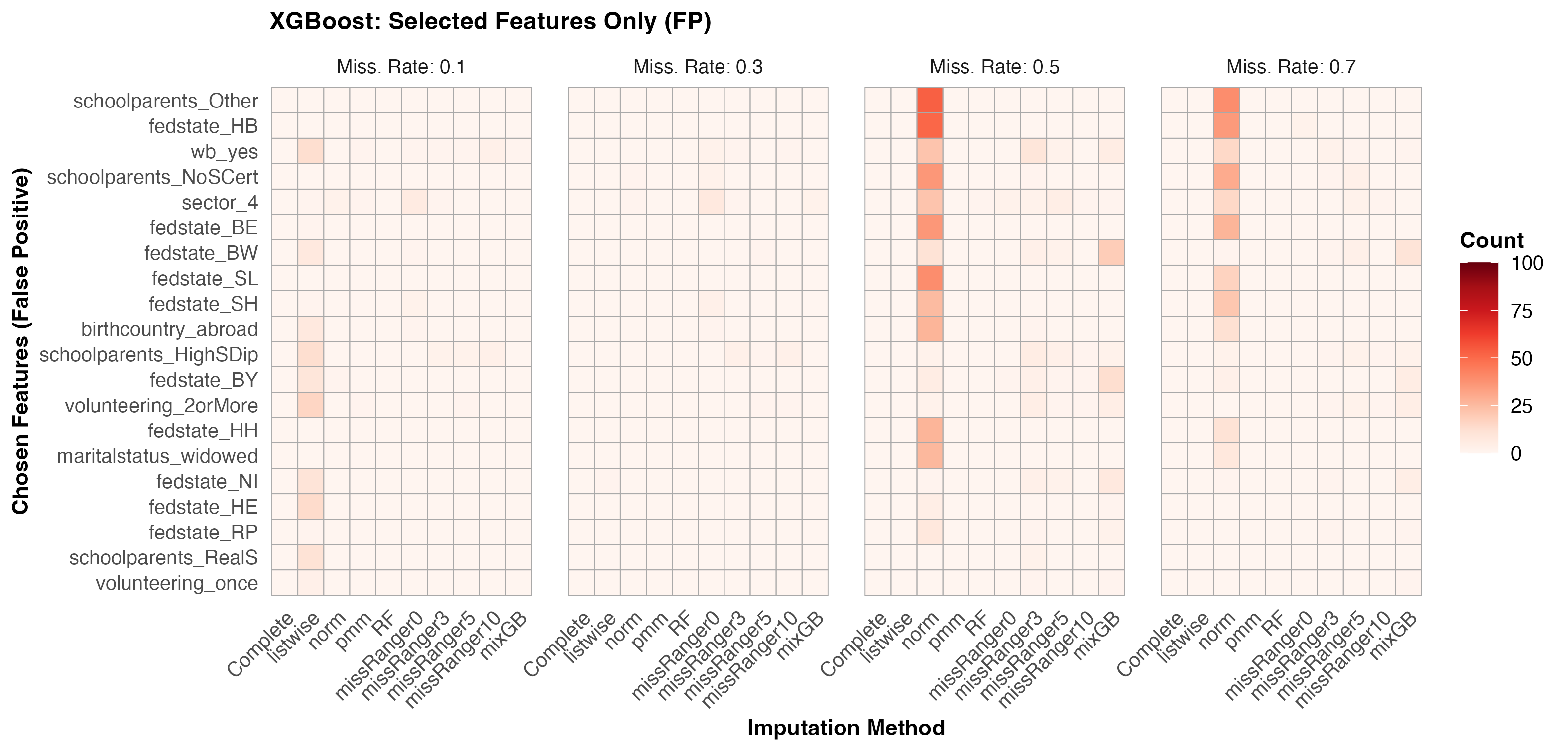}
\end{figure}

\textbf{False Positives:} Examining the false positives for the XGBoost model, as shown in Figure \ref{fig:featsXGBb}, it is apparent that \texttt{listwise deletion} is responsible for selecting more false features at the lowest missing rate compared to the baseline scenario. Conversely, at the higher missing rates, MICE \texttt{Norm}, together with the \texttt{missRanger} PMM versions and \texttt{mixGB}, are more prone to generating false positives.
In particular, at 50\% and 70\% missing rates, the largest number of false positive features occurs after imputation with MICE \texttt{Norm}. Here, the features \textit{schoolparents\_Other} and \textit{fedstate\_HB} are the most frequently misidentified, appearing in 53 and 51 cases at the 50\% missing rate, respectively. Thus, even though \texttt{MICE Norm} was best for selecting true positives with XGBoost, it is also the worst in terms of false positives when the missing rate is high.

\section{Discussion}

In this study, we evaluated how nine different imputation methods -- \texttt{listwise deletion}, \texttt{MICE Norm}, \texttt{MICE PMM}, \texttt{MICE RF}, \texttt{missRanger} without PMM, \texttt{missRanger} with 3, 5, or 10 donors, and \texttt{mixGB} -- affect the feature selection performance of three state-of-the art regression models: LASSO, Random Forest, and XGBoost. To this end, we used a subset of the NEPS Adult Cohort SC6 survey and examined how the feature selection of models trained on the full dataset compared to those trained on the imputed datasets.. Our results show that the choice of imputation method significantly affects the effectiveness of feature selection, with certain imputation-model pairings outperforming others under certain missing data conditions.

\subsubsection{LASSO}

When using LASSO,  \texttt{missRanger} without Predictive Mean Matching (PMM) consistently retained the highest number of true positive features across all missing rates. In fact, it was the method that best replicated the feature selection from the complete dataset compared to all other options. MICE Random Forest (MICE RF), \texttt{missRanger} with PMM, and \texttt{mixGB} also performed well, particularly at lower missing rates (10\% and 30\%). However, LASSO struggles to retain features when using listwise deletion and \texttt{MICE Norm}. LASSO inherently produced very few false positives across all imputation methods. Any observed false positives were minimal and primarily occurred with \texttt{missRanger} variants and \texttt{mixGB}, but these instances were negligible.

MissRanger without PMM offered the best balance for LASSO, maximizing true positives while virtually eliminating false positives. MICE RF and \texttt{mixGB} also provided a reasonable balance, though slightly less effective than \texttt{missRanger} without PMM.

\subsubsection{Random Forest}
Random Forests were robust in retaining true positives across all imputation methods, showing little sensitivity to the choice of imputation. Nearly half of the features selected in the complete data were consistently chosen, even at higher missing rates. However, Random Forests tended to select more false positives. In particular, \texttt{MICE Norm} and listwise deletion resulted in higher false positive rates, especially at lower missing rates (10\% and 30\%). At higher missing rates (50\% and 70\%), \texttt{MICE Norm} continued to produce more false positives.

No single imputation method stood out for Random Forests in balancing true and false positives. Avoiding \texttt{MICE Norm} and listwise deletion, however, reduced the number of false positives without compromising true positive selection.

\subsubsection{XGBoost}
\texttt{MICE Norm} facilitated the selection of more true positive features with XGBoost. However, it also led to a higher rate of false positives at increased missing rates. \texttt{missRanger} with PMM and \texttt{mixGB} introduced fewer false positives compared to \texttt{MICE Norm}.

MICE with PMM, MICE RF, \texttt{missRanger} with PMM, and \texttt{mixGB} provided a balanced approach with XGBoost, maintaining a reasonable number of true positives while minimizing false positives.

\subsubsection{Comparing the Machine Learning Methods}

Regarding feature selection, LASSO was the most conservative of the machine learning methods, yielding fewer true positives, particularly at higher missing rates. However, it also introduced the fewest false positives, making it ideal when minimizing false positives is essential. In contrast, Random Forests maintained the highest number of true positives across all imputation methods and missing rates, albeit with a higher count of false positives. XGBoost struck a balance, providing a moderate number of both true and false positives, particularly when paired with effective imputation methods like MICE with PMM or \texttt{mixGB}. This positioned XGBoost as a reasonable trade-off between true and false positives. However, combining XGBoost with MICE Norm at higher missing rates (50\% or 70\%) led to a significant rise in false positives, with the lowest false positive rates observed using MICE PMM or MICE RF.

\subsubsection{Relation to Previous Literature}

Our findings align with prior research emphasizing the importance of selecting suitable imputation methods in conjunction with machine learning models. The strong performance of tree-based imputation methods, such as \texttt{missRanger} without PMM, corroborates their effectiveness in handling complex, high-dimensional datasets with mixed data types \citep{Hayes2018, HayesM2017}. The increased false positives associated with MICE Norm, particularly with Random Forest and XGBoost at higher missing rates, support concerns about the limitations of linear model-based imputations in datasets with nonlinear relationships \citep{thurow2021imputing}. Compared to \cite{SchroederSKD2024}, who examined the impact of imputation methods on the use of tree-based prediction rule ensembles \citep{FokkemaS2020}, the differences between tree-based imputation methods in our study are small.

As is typically observed in the literature, listwise deletion yielded inferior results \citep{VanGinkelEA2020,Pepinsky_2018, myers2011goodbye}. It demonstrated consistent underperformance across all combinations of imputation models, missing rates, and ML models, yielding to 
misaligned feature selections in comparison to the baseline cases.

In the context of feature selection, the application of PMM to \texttt{missRanger} resulted in the selection of a greater number of false positive features than in its absence. Even though \texttt{missRanger} with PMM might be better suited for statistical inference \citep{schwerter2024evaluating}, it might be worse for feature selection.

\subsubsection{Recommendations}

Our results add to the body of research that recommends avoiding listwise deletion. Furthermore, MICE standards for imputing missing data should be avoided when the primary goal is feature selection, as these methods tend to underperform in preserving important features and may lead to higher false positive rates. To achieve a sparse model with minimal false positives, LASSO is recommended, especially when paired with tree-based imputation methods. XGBoost provides a balanced approach between true positives and false positives, making it suitable for broader variable selection. The choice of feature selection method should be based on the research objectives: LASSO is ideal for identifying a sparse set of variables, while Random Forest selects a larger number of features, with XGBoost positioned in between. However, researchers should note that an increase in true positives often correlates with an increase in false positives, with LASSO generally being more effective than XGBoost, which in turn outperforms Random Forest in minimizing false positives. In addition, Random Forest and XGBoost are more resilient to missing data than LASSO when missingness exceeds 30\%. Therefore, using XGBoost as a statistical learner after multiple imputation can achieve a good balance with optimal mean squared error results, while the right imputation method will further help reduce false positives.

\subsection{Conclusions}\label{sec:conclusion}

Identifying a universally optimal combination of imputation and machine learning models for feature selection tasks is challenging. Practitioners must balance the importance of retaining true positives against the risk of introducing false positives. In our study, XGBoost also demonstrated superior Mean Squared Error (MSE) performance (see Appendix \ref{appendix:addResults}), paired with imputation methods like MICE with PMM or RF and \texttt{mixGB}. It thus offers a pragmatic balance. Ultimately, the choice should align with the specific goals of the research, the nature of the data, and acceptable levels of false positives within the study’s context. Furthermore, other considerations, such as interpretability or the ability to quantify uncertainty, could influence this recommendation and should also be factored into the decision.


\subsection{Outlook}

To generalize our results and more thoroughly evaluate imputation models, future studies should use datasets with different parameters and observations and include different variable types (numeric, categorical, integer). In addition, tuning hyperparameters for imputation methods such as Random Forests or XGBoost could improve their results and help practitioners identify the most relevant features during the imputation process.

A deeper investigation of threshold selection for feature filtering after training Random Forests or XGBoost could determine whether these methods can match or exceed the performance of LASSO in terms of false positives. In addition, examining the effects of retraining these models using the most significant features from the initial filtering could provide valuable insights.

Finally, while many false positives were associated with dummy variables corresponding to true positive features, future research could explore whether broader hot-encoding strategies -- such as grouping similar states -- could reduce the number of input features without sacrificing explanatory power, resulting in fewer false positives.

Finally, while our study found that \texttt{missRanger} without PMM performed better for feature selection, it underperformed in statistical inference in other studies \citep{GurtskaiaSD2024, schwerter2024evaluating}. Conversely, MICE Norm, while not effective for feature selection in our study, has demonstrated strong performance in official statistics, particularly in terms of univariate and multivariate distributional imputation accuracy for metric variables \citep{thurow2021imputing,thurow2024assessing}, and provided reliable type I error control after imputation \citep{ramosaj2020cautionary}. This raises the question of whether a single study, especially one aimed at selecting features for post-selection regressions \citep{BelloniCH2013, SchwerterDBM2022}, should use different imputation methods for feature selection and regression analysis, at least when traditional distributional assumptions (like normality) are not met.

\section*{Acknowledgments}

\emph{This document uses data from the National Educational Panel Study (\cite{neps_network_2022}; see \cite{blossfeld2019education}). The NEPS is carried out by the Leibniz Institute for Educational Trajectories (LIfBi, Germany) in cooperation with a nationwide network.
}

\emph{The authors gratefully acknowledge the computing time provided on the Linux HPC cluster at Technical University Dortmund (LiDO3), partially funded as part of the Large-Scale Equipment Initiative by the Deutsche Forschungsgemeinschaft (DFG, German Research Foundation) as project 271512359.}

\emph{Moreover, this work has been partly supported by the Research Center for Trustworthy Data Science and Security (https://rc-trust.ai), one of the Research Alliance centers within the https://uaruhr.de.}

\section*{Declarations}

\subsection*{Open Practice}

Materials and analysis code for the simulation study are available at \href{}{?}.

\subsection*{Funding}

The project ``From Prediction to Agile Interventions in the Social Sciences (FAIR)'' has received funding from the programme ``Profilbildung 2020'' (PROFILNRW-2020-068), an initiative of the Ministry of Culture and Science of the State of North Rhine Westphalia. Sole responsibility for the content of this publication lies with the authors.

The authors gratefully acknowledge the computing time provided on the Linux HPC cluster at Technical University Dortmund (LiDO3), partially funded as part of the Large-Scale Equipment Initiative by the Deutsche Forschungsgemeinschaft (DFG, German Research Foundation) as project 271512359.

\subsection*{Conflict of interest}

There is no conflict of interest

\subsection*{Authors contribution}
Author contributions: \insertcreditsstatement


\bibliographystyle{tfcse}
\bibliography{literature}
\newpage
\appendix

\section{Mappings and Detailed Descriptions for Factor Variables} 
\label{appendix:factors}

Along this section are listed the different levels for each feature in frequency descending order.

\subsection{Classification of Profession Requirement}
The requirement for the profession of the respondent according to the KldB (in German: Klassifikation der Berufe).
\begin{table}[htbp!]
	\caption{Levels Available in \texttt{KLDB}.} \label{tab:kldb}
	\centering
	\small
	\begin{tabular}{cl}
		\toprule
		\textbf{Final Level} & \textbf{Description} \\
		\midrule
		\textit{Skilled} & Specialized activities \\
		\textit{HighComplx} & Highly complex activities \\
		\textit{Complx} & Complex specialist activities \\
		\textit{LowComplx} & Helpers and semi-skilled work \\
		\bottomrule
	\end{tabular}
\end{table}

\subsection{Marital Status}

\begin{table}[htbp!]
\caption{Levels Available in \texttt{maritalstatus}.} \label{tab:maritalstatus}
\centering
\small
\begin{tabular}{cl}
	\toprule
	\textbf{Final Level} & \textbf{Description} \\
	\midrule
	\textit{married} & Married or in registered civil partnership \\
	\textit{single} & Single \\
	\textit{divorced} & Divorced \\
	\textit{widowed} & Widowed \\
	\bottomrule
\end{tabular}
\end{table}

\subsection{Educational Level}

For this feature, eight levels were initially  available but they were reduced to just 4 in order to reduce the dimensionality of the data. In Table \ref{tab:education} are shown the mappings made and the former levels with their descriptions.

\begin{table}[H]
	\caption{Mappings and Detailed Description for \texttt{education}.} \label{tab:education}
	\centering
	\small
	\begin{tabular}{cl}
		\toprule
		\textbf{Final Level} & \textbf{Former Level / Description} \\
		\midrule
		\multirow{3}{*}{\textit{LowerSec}} & no qualification \\
		 & Lower secondary school leaving certificate without vocational training \\
		 & Lower secondary school leaving certificate with vocational training \\
		  & \\[-7pt]
		\multirow{2}{*}{\textit{Secondary}}& Secondary school leaving certificate without vocational training \\
		 & Secondary school leaving certificate with vocational training \\
 		  & \\[-7pt]
		\multirow{2}{*}{\textit{Abitur}} & Abitur without vocational training \\
		 & Abitur with vocational training \\
		  & \\[-7pt]		 
		\multirow{2}{*}{\textit{University}} & University of Applied Sciences degree \\
	 	 & University degree \\
		\bottomrule
	\end{tabular}
\end{table}

\vskip 3cm

\subsection{Company Size}
The size of the company in terms of number of employees.
\begin{table}[H]
	\caption{Levels Available in \texttt{compsize}.} \label{tab:compsize}
	\centering
	\small
	\begin{tabular}{cl}
		\toprule
		\textbf{Final Level} & \textbf{Description} \\
		\midrule
		3 & 20 to less than 200 persons \\
		4 & 200 persons or more \\
		1 & Less than 10 persons \\
		2 & 10 to less than 20 persons \\
		\bottomrule
	\end{tabular}
\end{table}

\subsection{Sector}
The industry sector of the company in which the respondent is working.
\begin{table}[H]
	\caption{Levels Available in \texttt{sector}.} \label{tab:sector}
	\centering
	\small
	\begin{tabular}{cl}
		\toprule
		\textbf{Final Level} & \textbf{Description} \\
		\midrule
		3 & Public and personal services \\
		2 & Economic services \\
		1 & Manufacturing \\
		4 & Other industries \\
		\bottomrule
	\end{tabular}
\end{table}

\subsection{Parents' Qualification}

Highest school-leaving qualification from surveyed's parents.
\begin{table}[H]
	\caption{Mappings and Detailed Description for \texttt{schoolparents}.} \label{tab:schoolparents}
	\centering
	\small
	\begin{tabular}{cl}
		\toprule
		\textbf{Final Level} & \textbf{Former Level / Description} \\
		\midrule
		\textit{HauptS} & Secondary school leaving certificate I $\rightarrow$ Haupt-, Volksschulabschluss \\
		\textit{HighSDip} & Secondary school leaving certificate III $\rightarrow$ (Technical) Abitur  \\
		\textit{RealS} & Secondary school leaving certificate II $\rightarrow$ Mittlere Reife, Realschulabschluss \\
		\textit{NoSCert} & No school-leaving qualification \\
		\textit{Other} & Other school-leaving qualification \\
		\bottomrule
	\end{tabular}
\end{table}

\subsection{Federal States}

Mapping according to the code for each German State.
\begin{table}[H]
	\caption{Mappings and Detailed Description for \texttt{fedstate}.} \label{tab:fedstate}
	\centering
	\small
	\begin{tabular}{cl}
		\toprule
		\textbf{Final Level } & \textbf{Former Level / Description} \\
		\midrule
		\textit{NW} & Nordrhein-Westfalen \\ 
		\textit{BY} & Bayern \\ 
		\textit{BW} & Baden-Württemberg \\
		\textit{NI} & Niedersachsen \\ 
		\textit{HE} & Hessen \\ 
		\textit{RP} & Rheinland-Pfalz \\ 
		\textit{SN} & Sachsen \\ 
		\textit{BE} & Berlin (Gesamt) \\ 
		\textit{ST} & Sachsen-Anhalt \\
		\textit{TH} & Thüringen \\ 
		\textit{BB} & Brandenburg \\ 
		\textit{SH} & Schleswig-Holstein \\ 
		\textit{HH} & Hamburg \\ 
		\textit{MV} & Mecklenburg-Vorpommern \\ 
		\textit{SL} & Saarland \\ \textit{HB} & Bremen \\ 
		\bottomrule
	\end{tabular}
\end{table}

\newpage
\section{Distribution Plots}
 \label{appendix:distplots}
In this section are plotted the histogram or count plots for each of the variables present in the dataset before they were encoded.

\subsection{Numeric Variables}
\begin{figure}[ht!]
	\caption{Distribution Plots for the Metric Variables}
	\begin{subfigure}{\textwidth}
		\centering
		\includegraphics[width=0.475\textwidth]{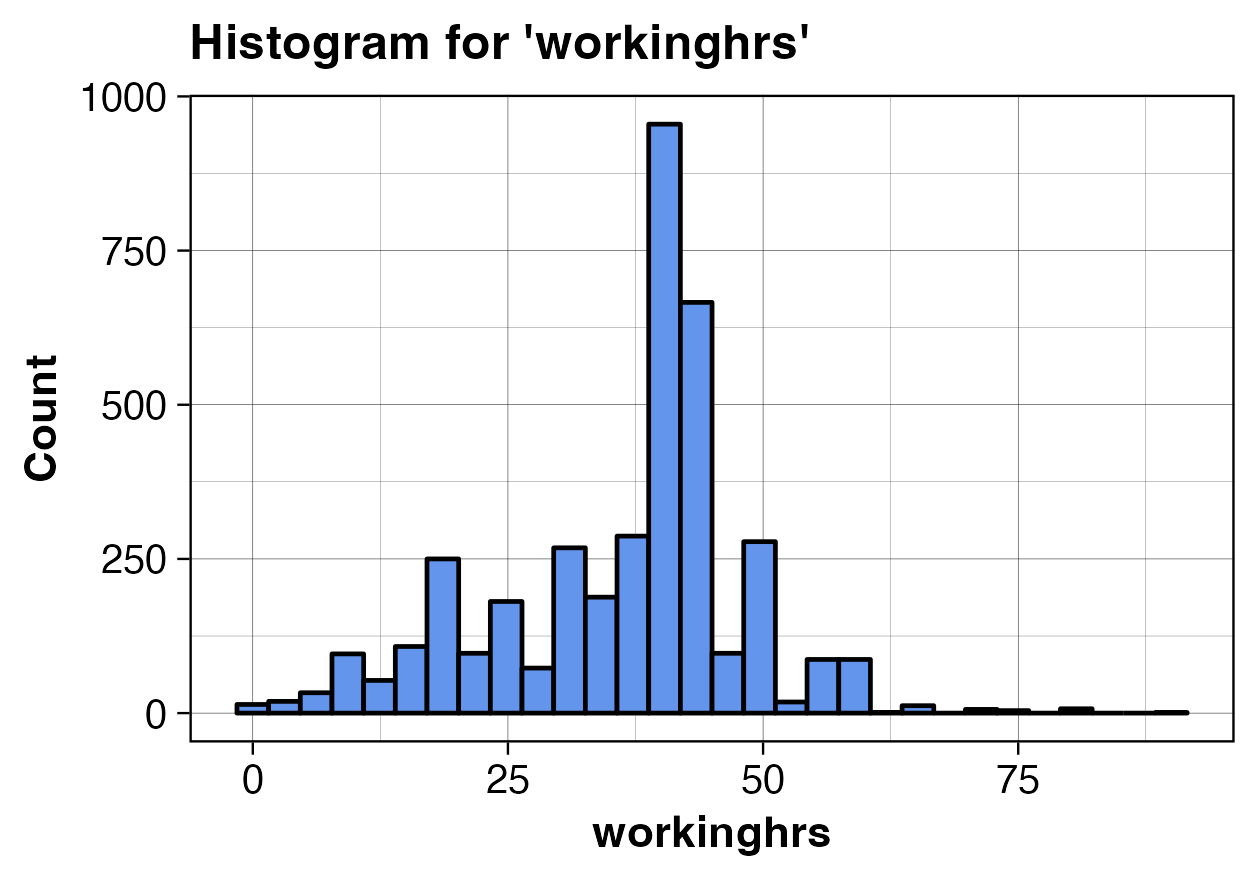} \hfill
		\includegraphics[width=0.475\textwidth]{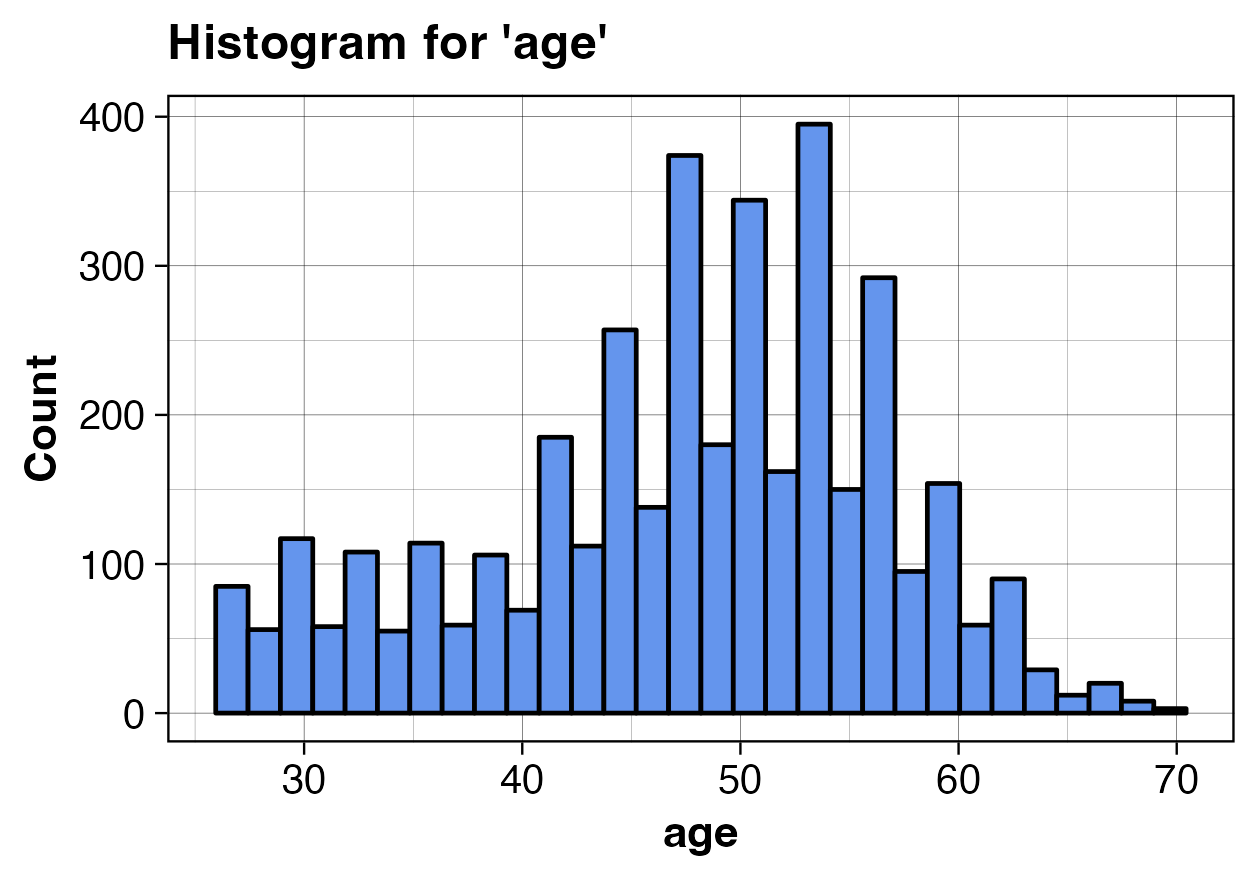} \\
		\includegraphics[width=0.475\textwidth]{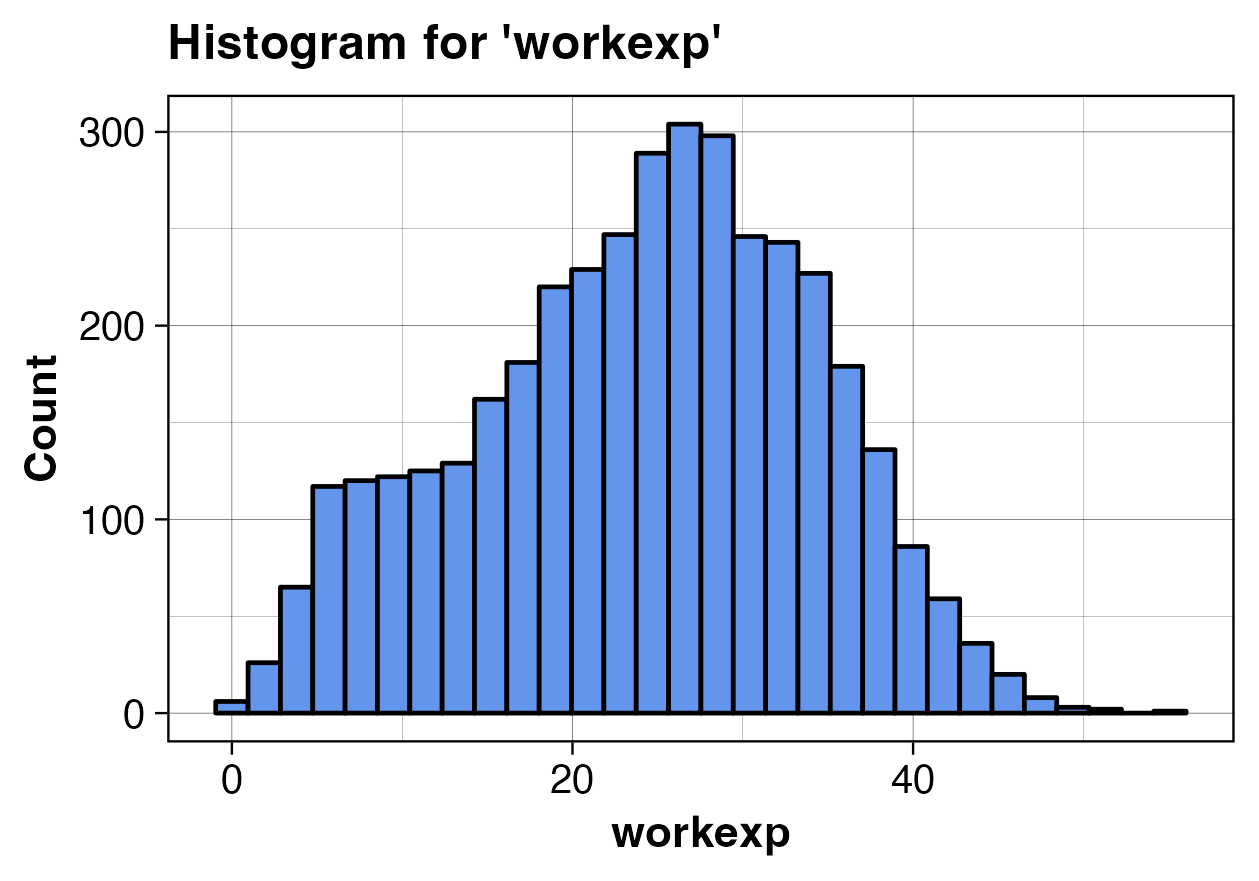} \hfill
		\includegraphics[width=0.475\textwidth]{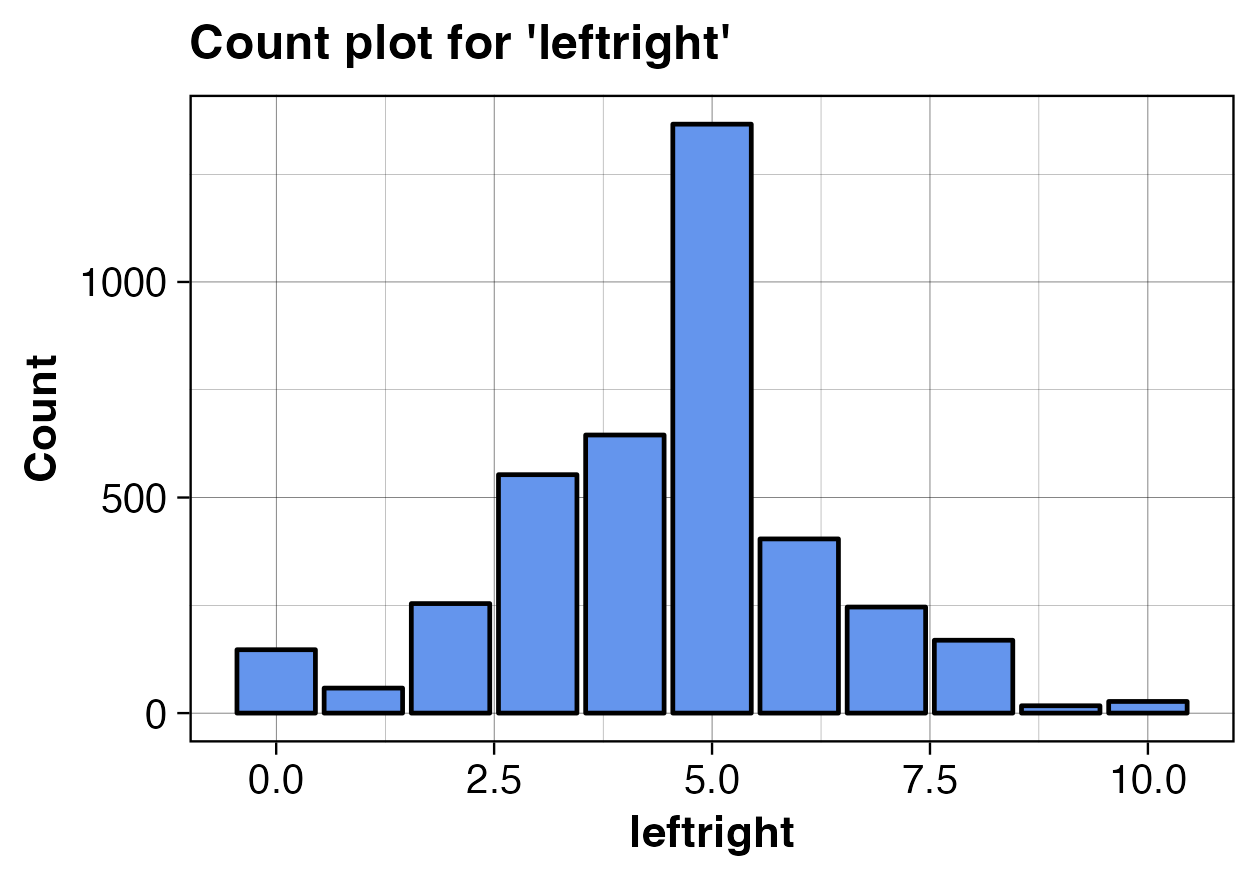} \\
	\end{subfigure}
\end{figure} 

\newpage

\subsection{Integer Variables}
\begin{figure}[ht!]
	\caption{Distribution Plots for the Integer Variables}
	\begin{subfigure}{\textwidth}
		\centering
		\includegraphics[width=0.475\textwidth]{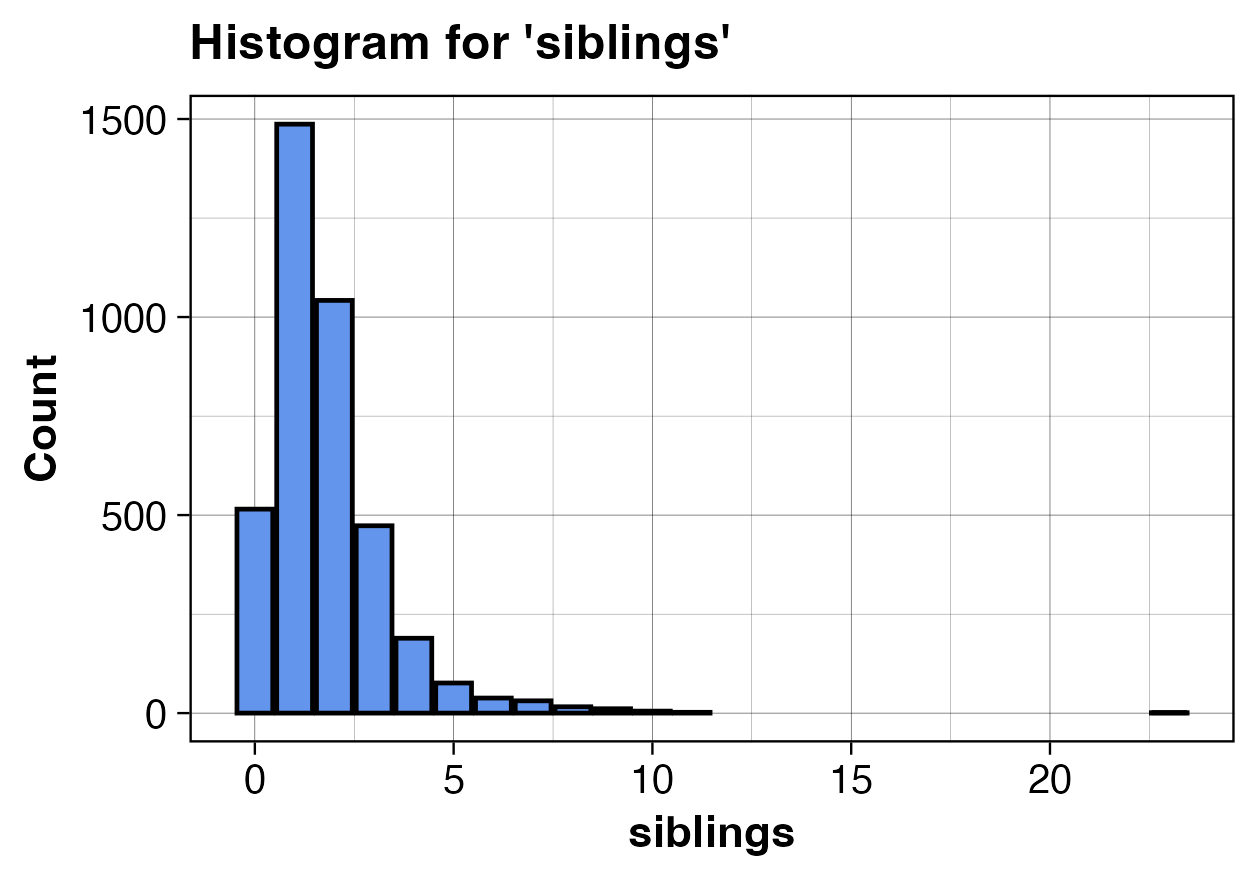} \hfill
		\includegraphics[width=0.475\textwidth]{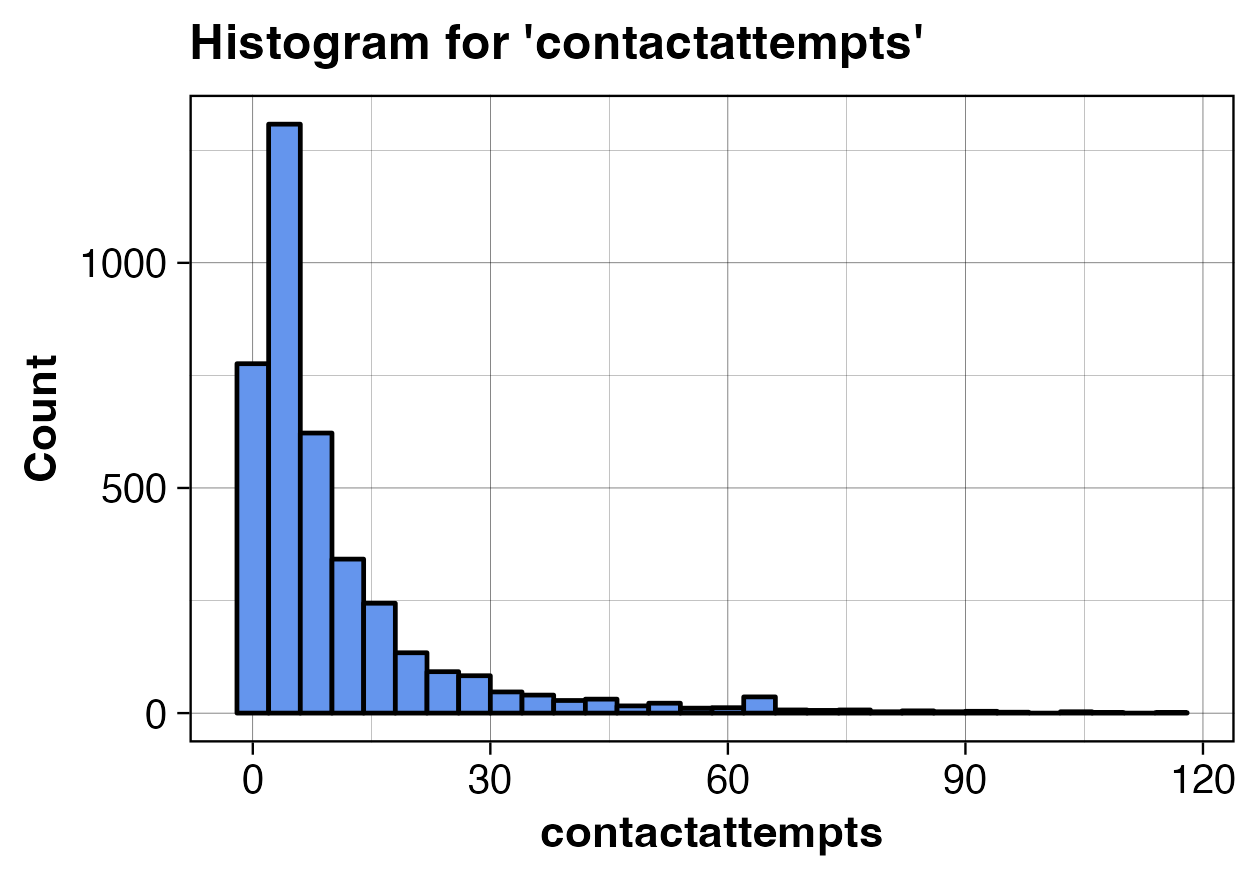} \\
        \includegraphics[width=0.475\textwidth]{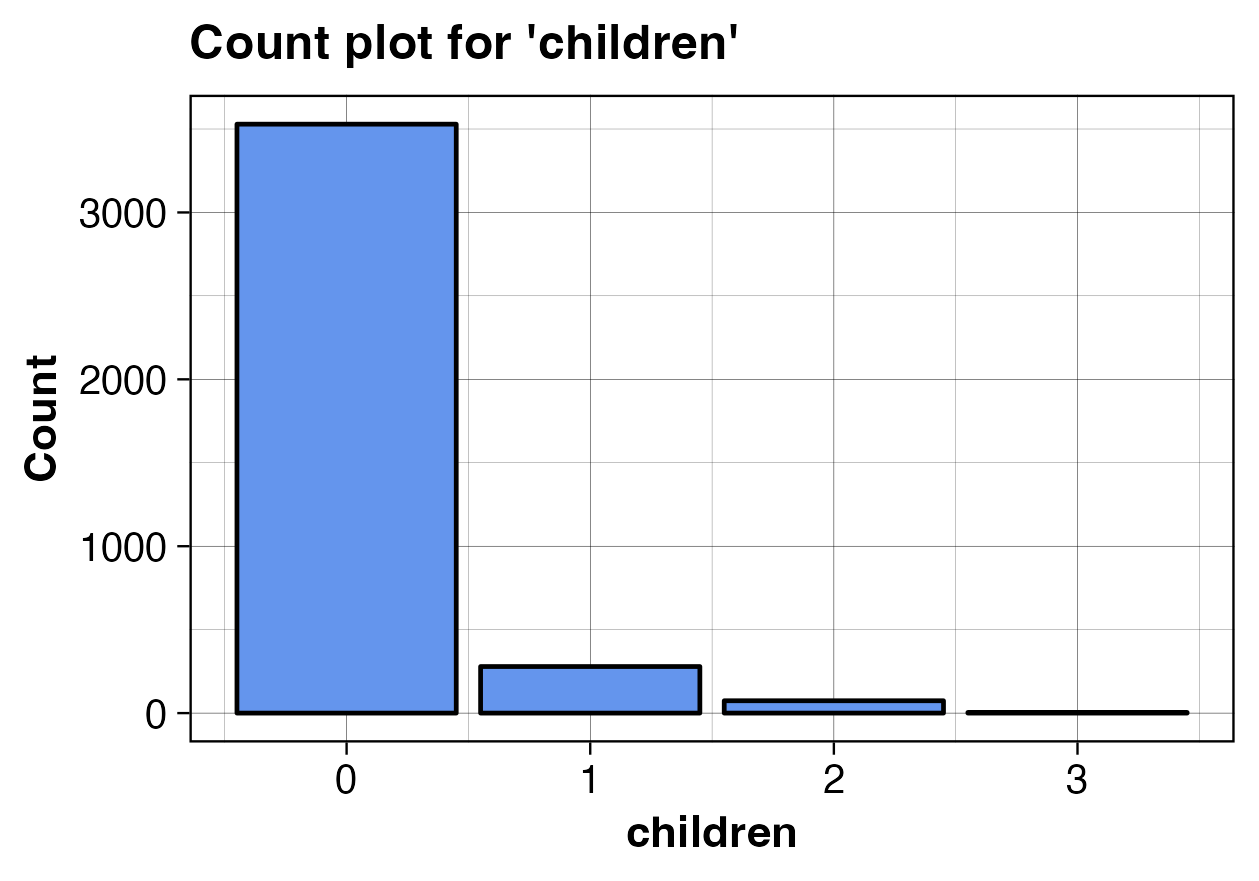} \hfill
	\end{subfigure}
\end{figure}

\subsection{Categorical Variables}
\begin{figure}[H]
	\caption{Distribution Plots for the Categorical Variables}
	\begin{subfigure}{\textwidth}
		\centering
		\includegraphics[width=0.475\textwidth]{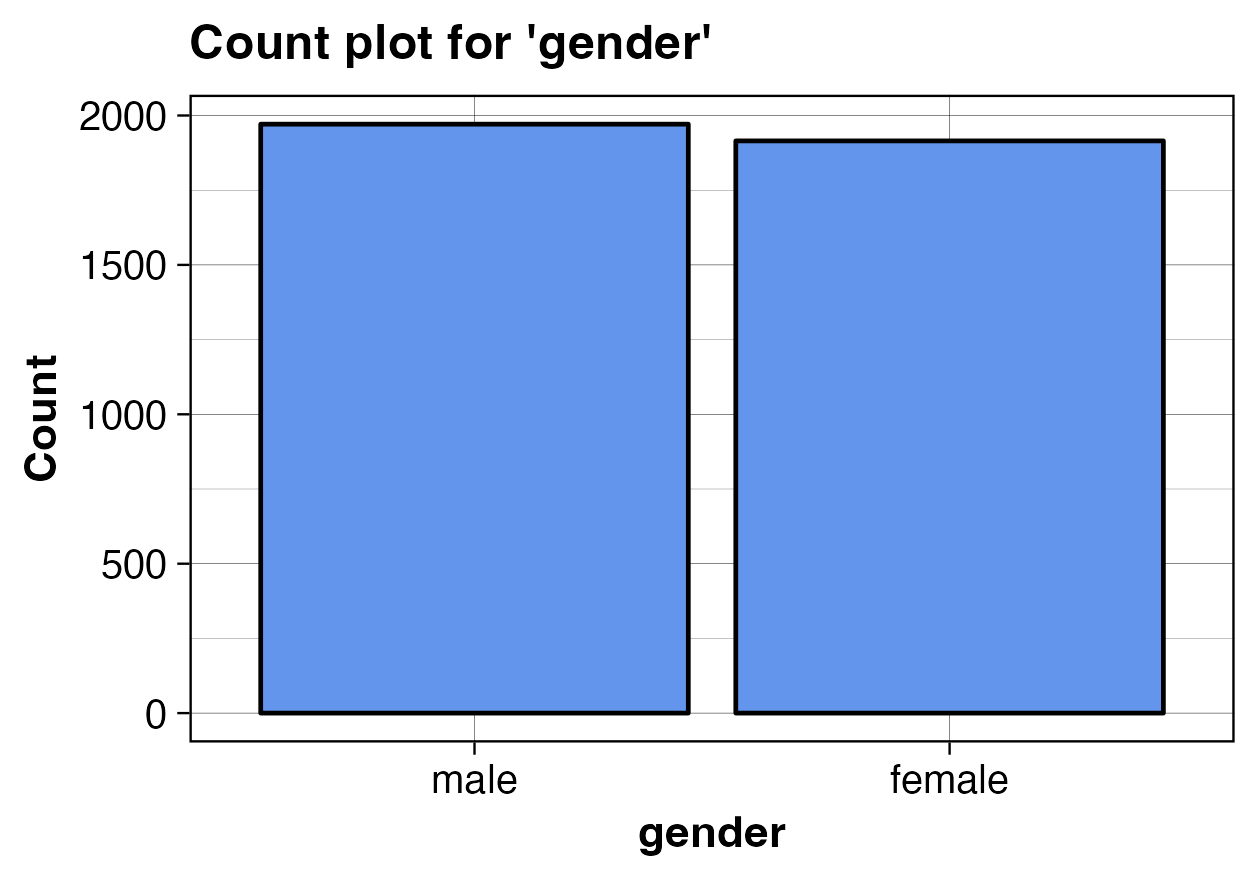} \hfill
		\includegraphics[width=0.475\textwidth]{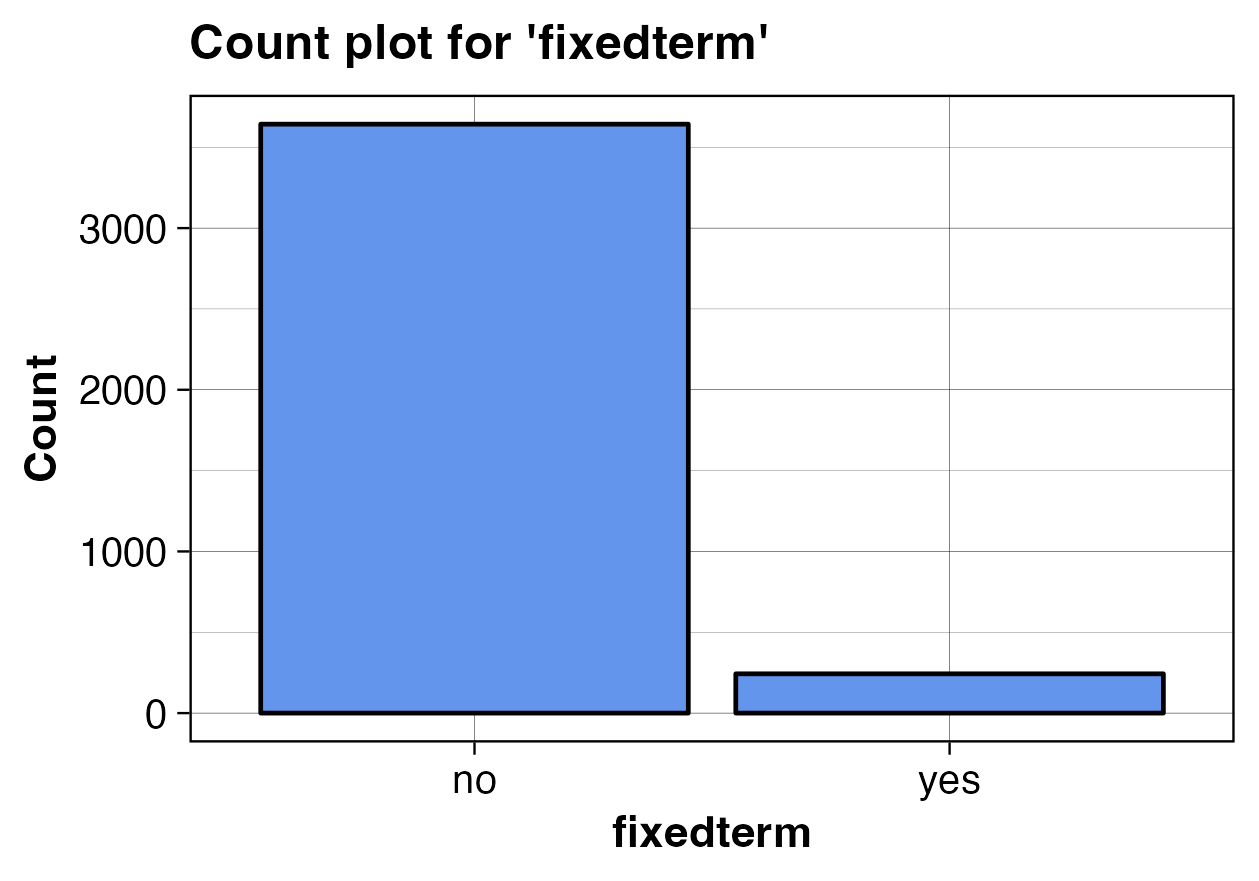} \\
		\includegraphics[width=0.475\textwidth]{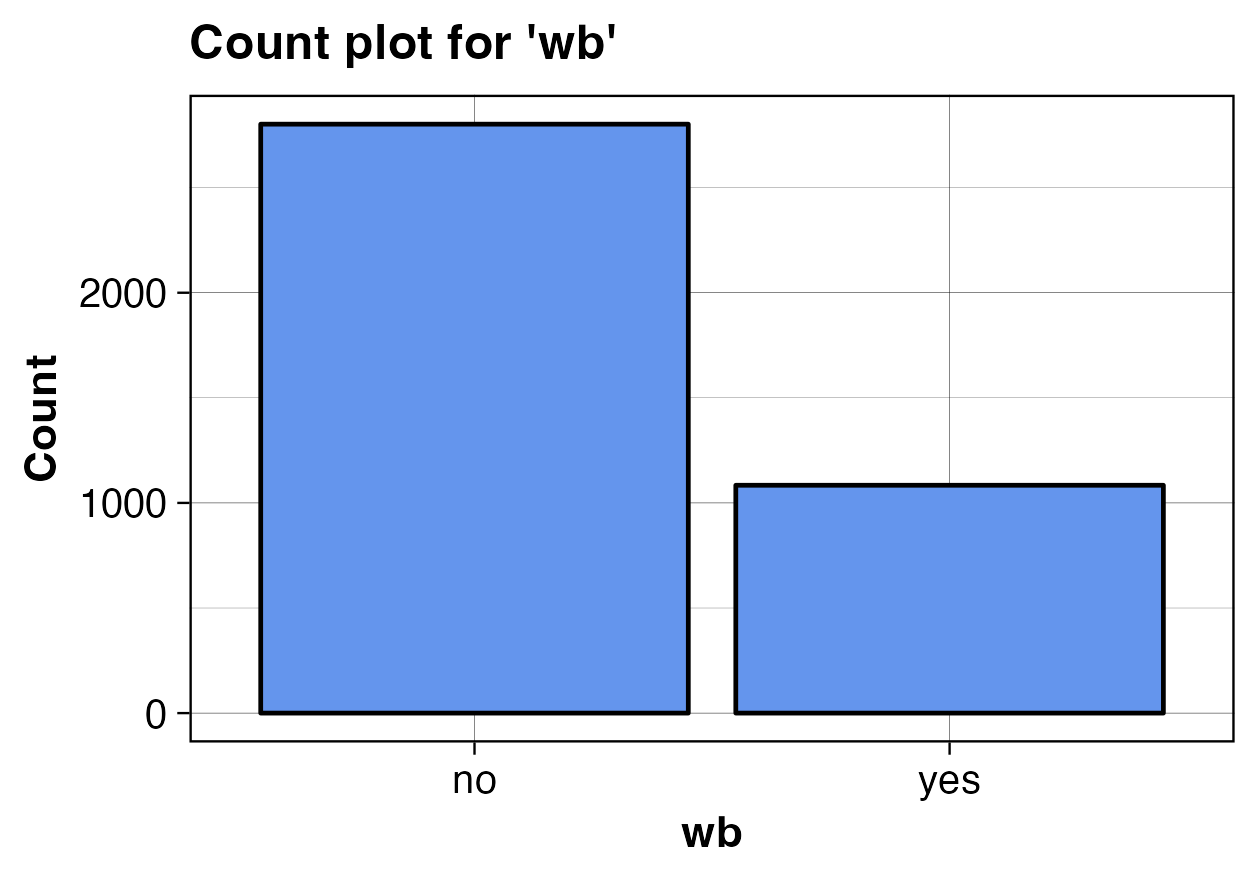} \hfill
		\includegraphics[width=0.475\textwidth]{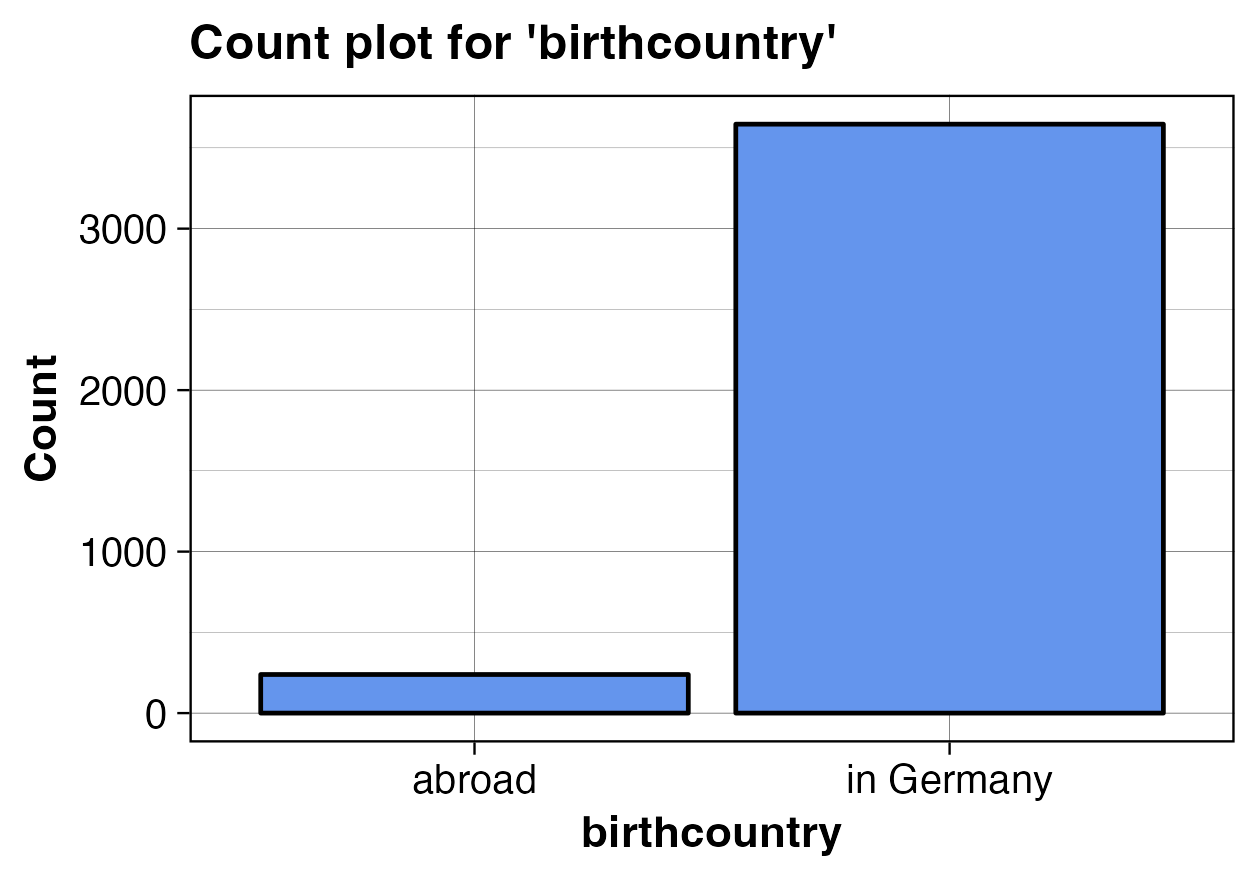} \\
        \includegraphics[width=0.475\textwidth]{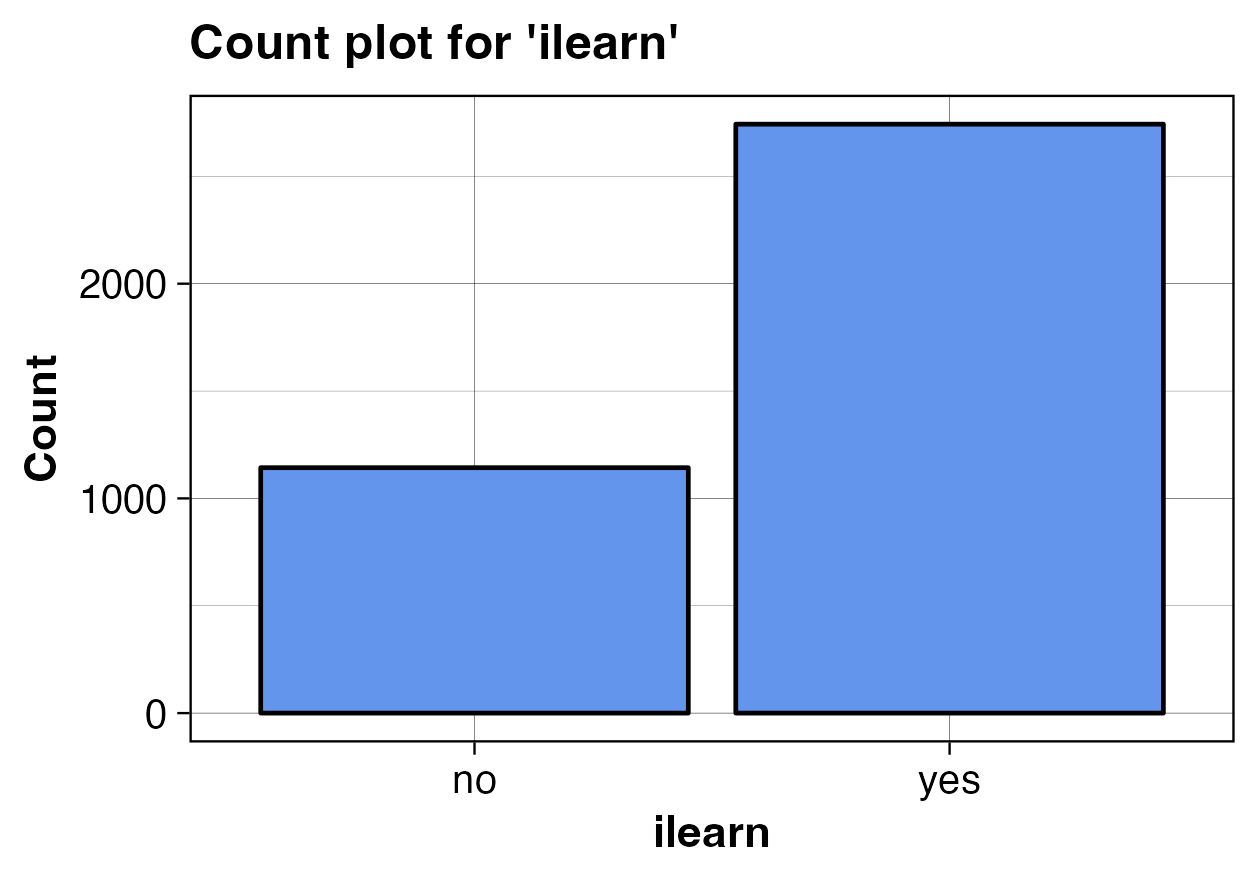} \hfill
		\includegraphics[width=0.475\textwidth]{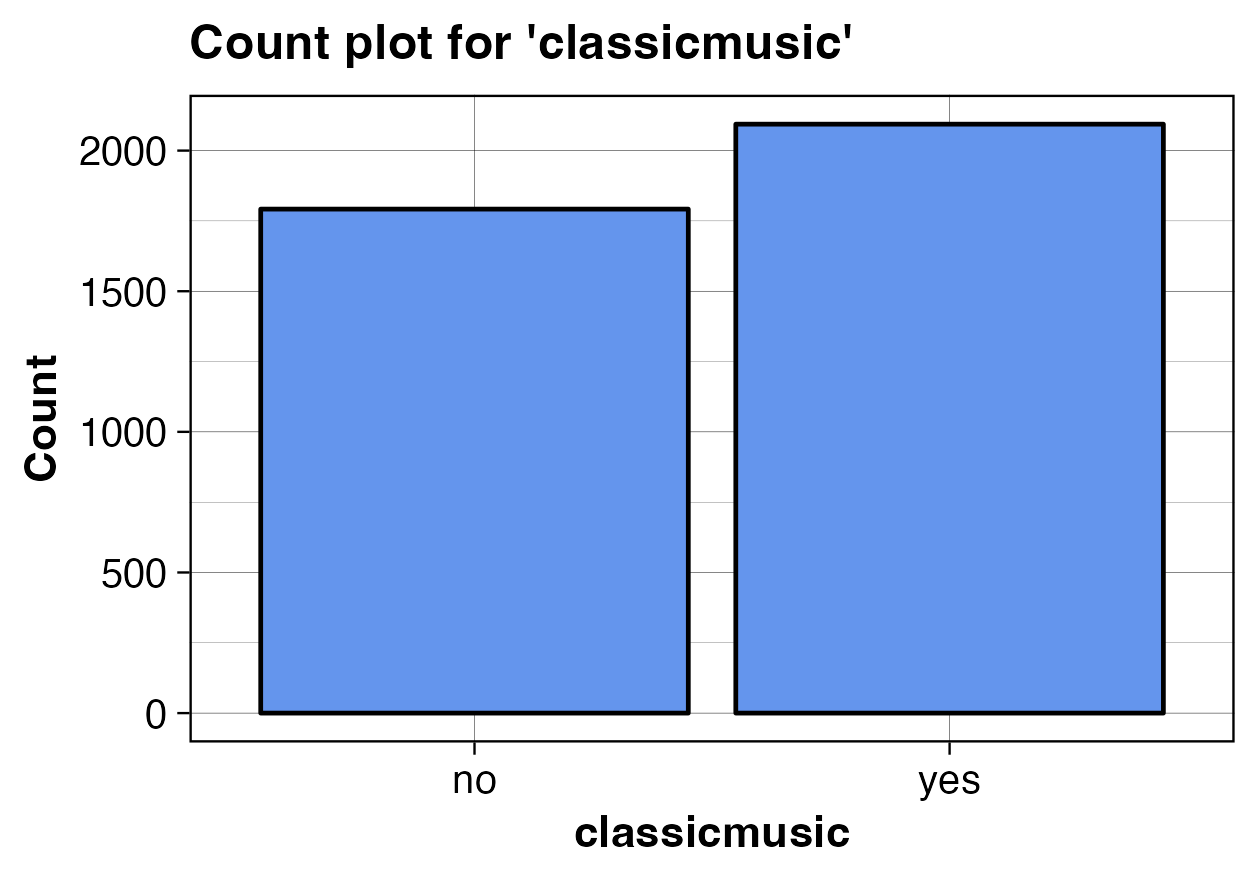} \\
		\includegraphics[width=0.475\textwidth]{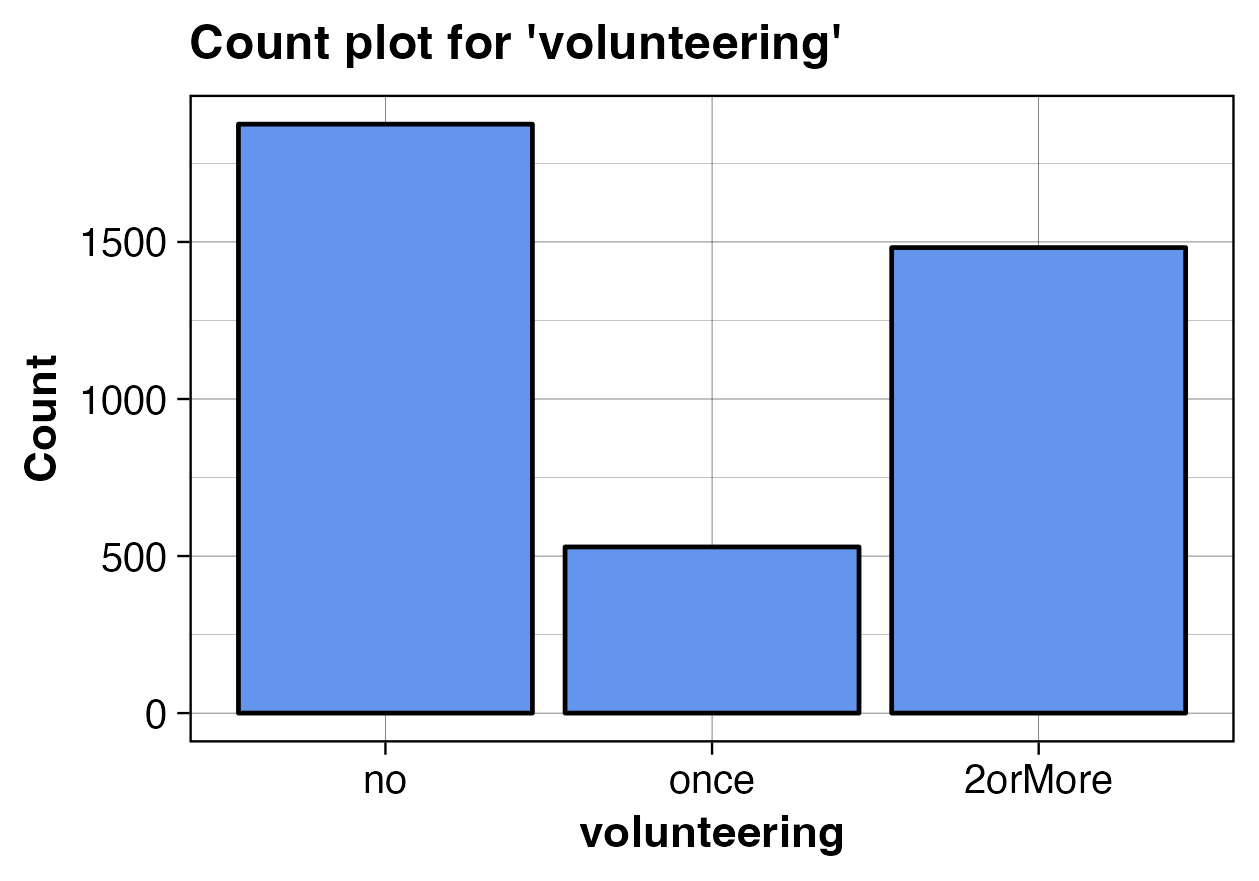} \hfill
		\includegraphics[width=0.475\textwidth]{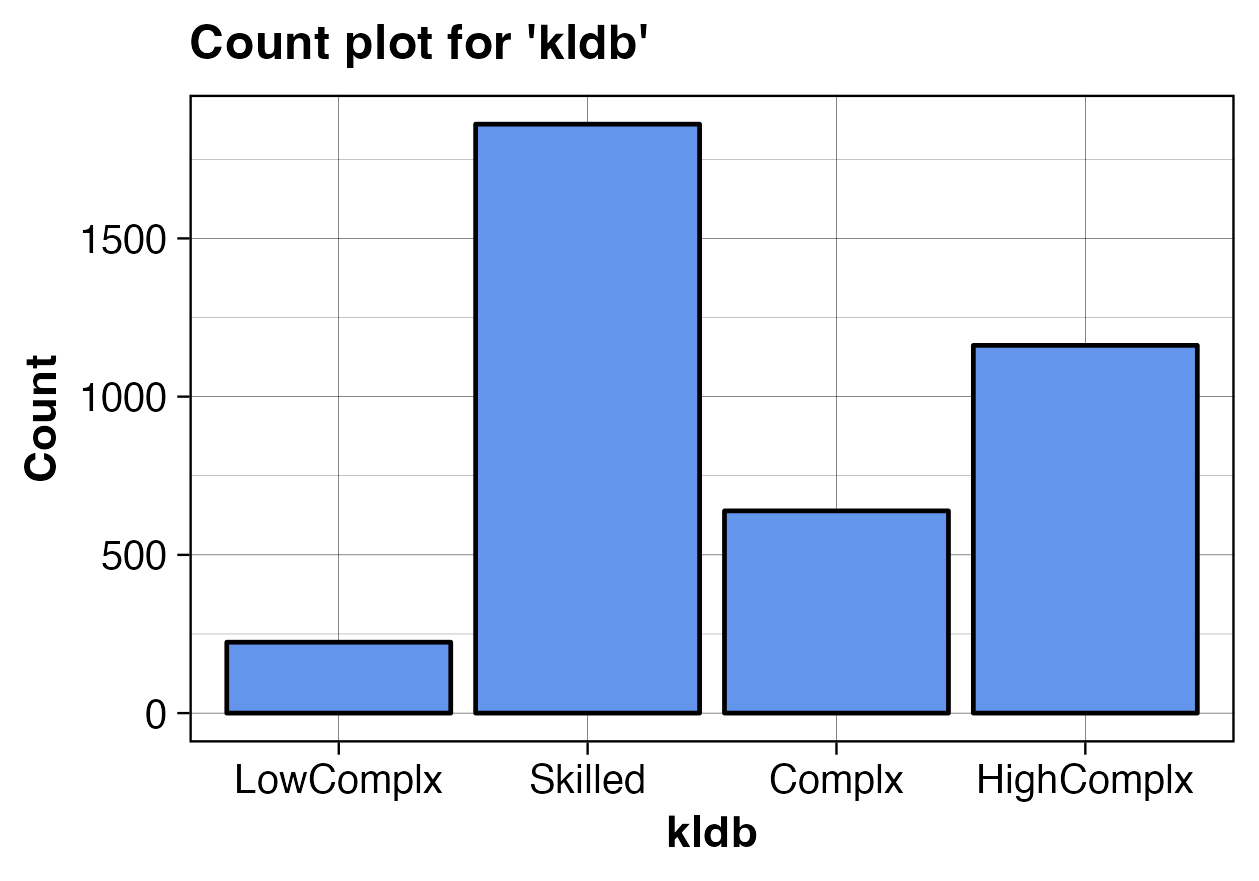} \\
	\end{subfigure}
\end{figure} 

\begin{figure}[H]
\begin{subfigure}{\textwidth}
		\includegraphics[width=0.475\textwidth]{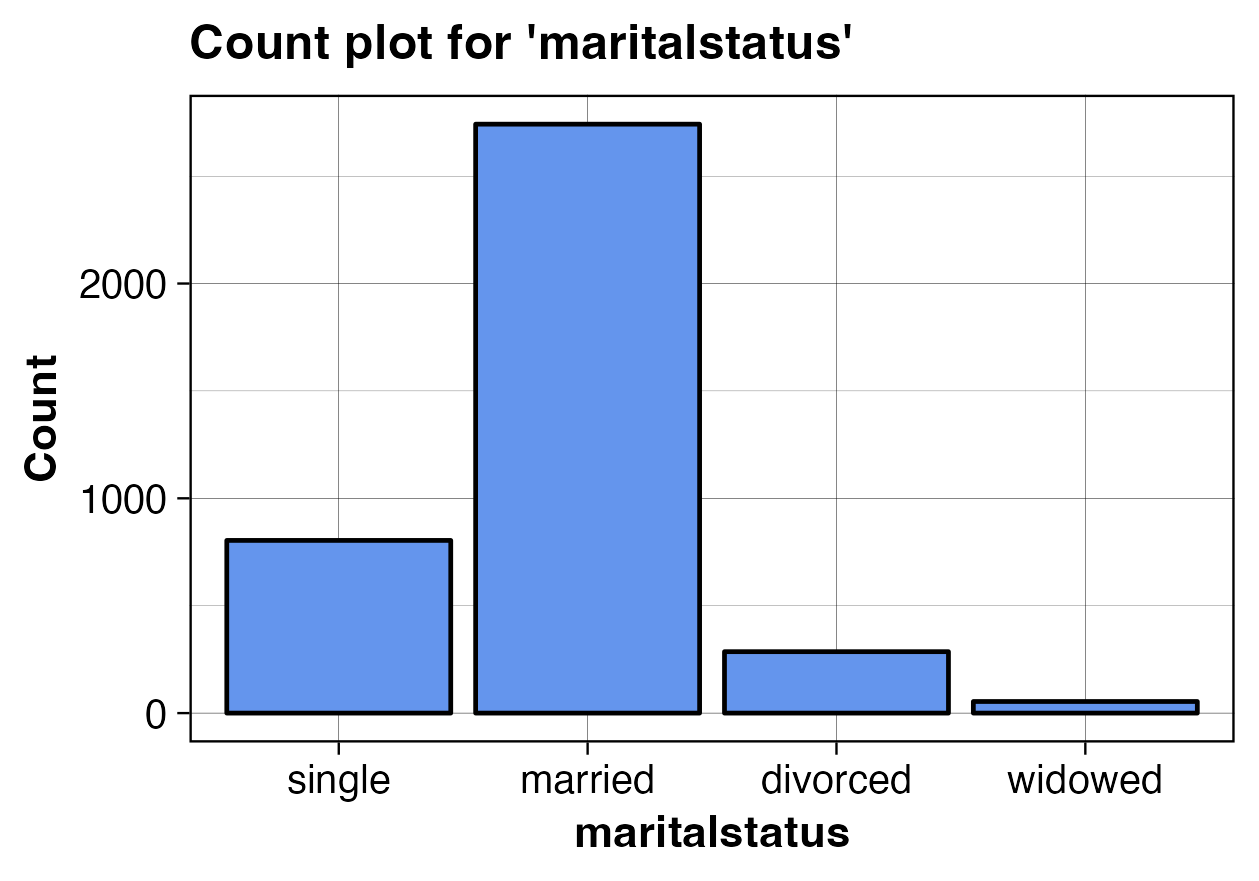} \hfill
		\includegraphics[width=0.475\textwidth]{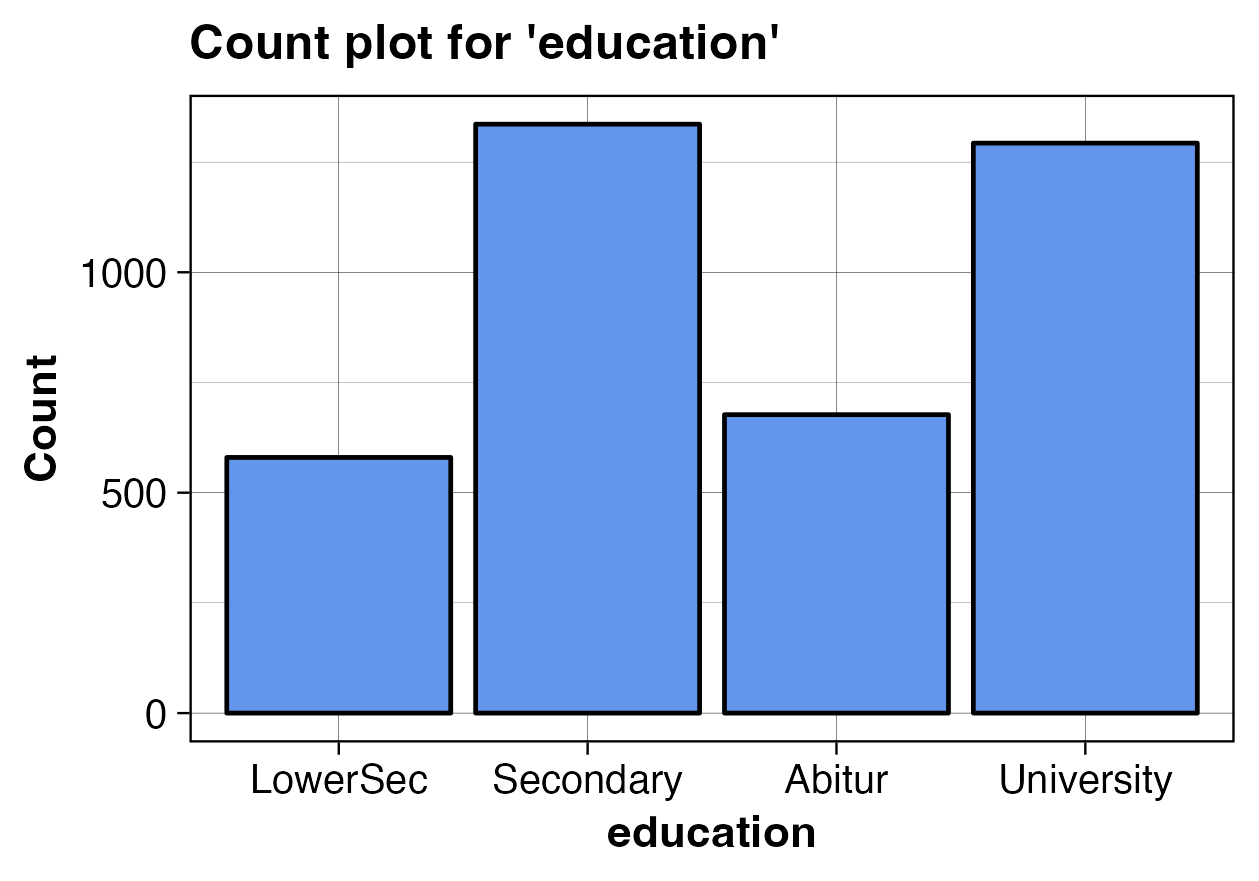} \\
		\includegraphics[width=0.475\textwidth]{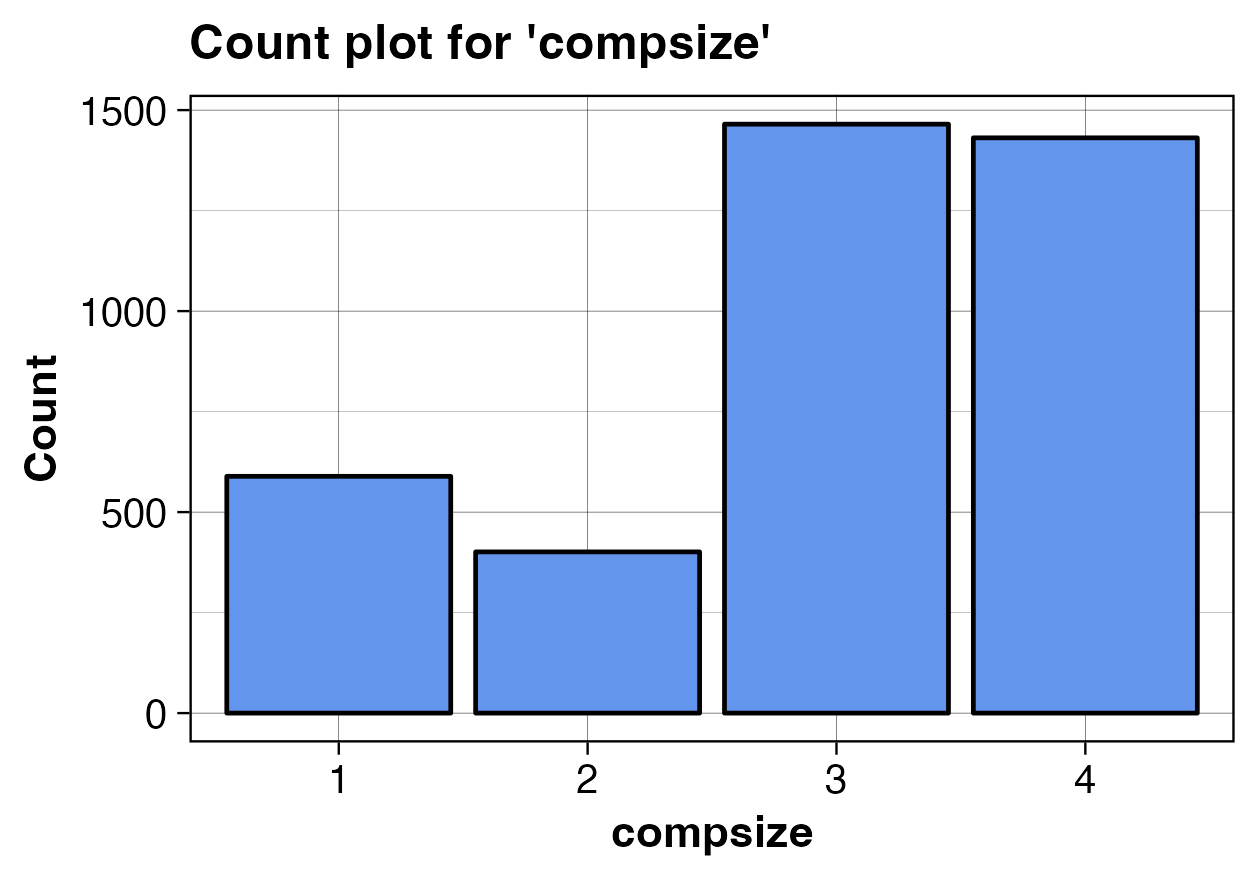} \hfill
		\includegraphics[width=0.475\textwidth]{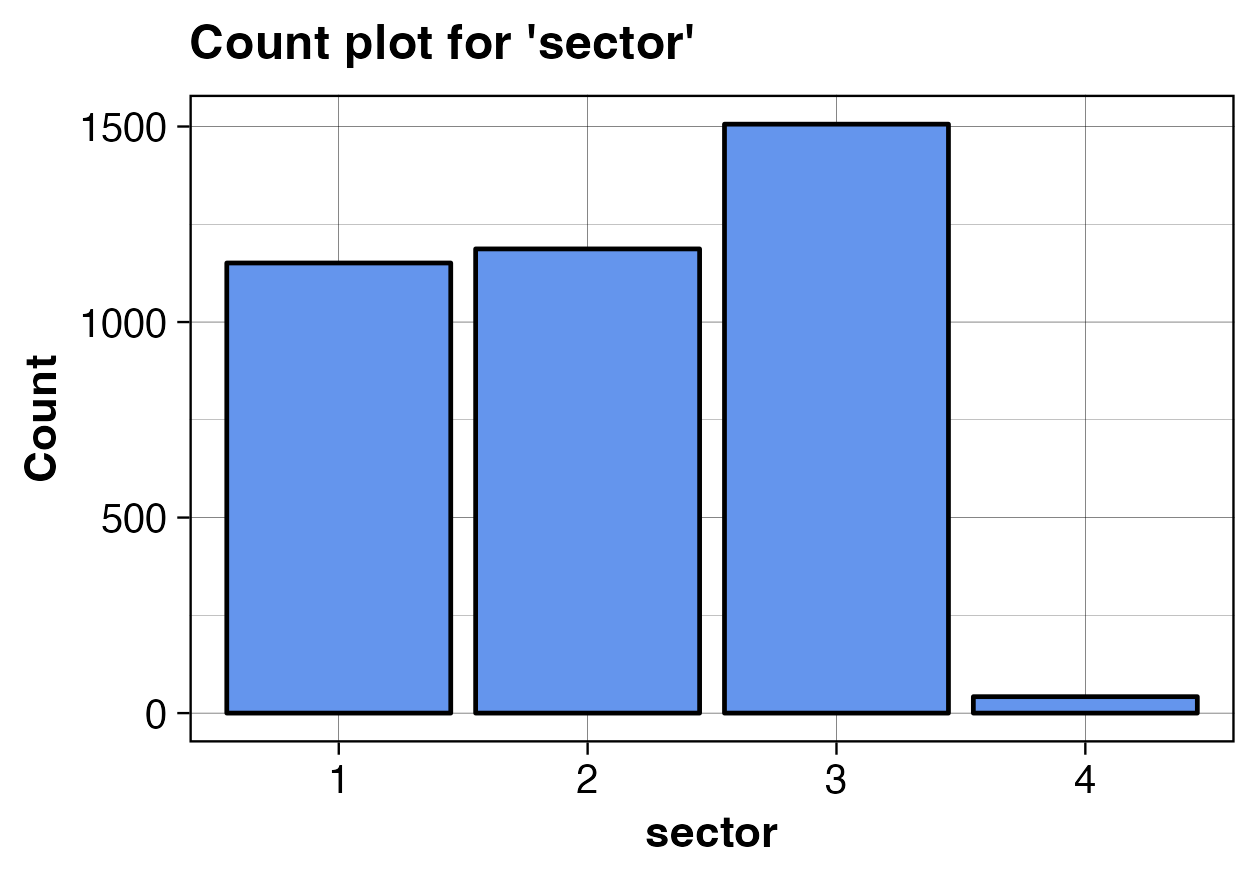} \\
        \includegraphics[width=0.475\textwidth]{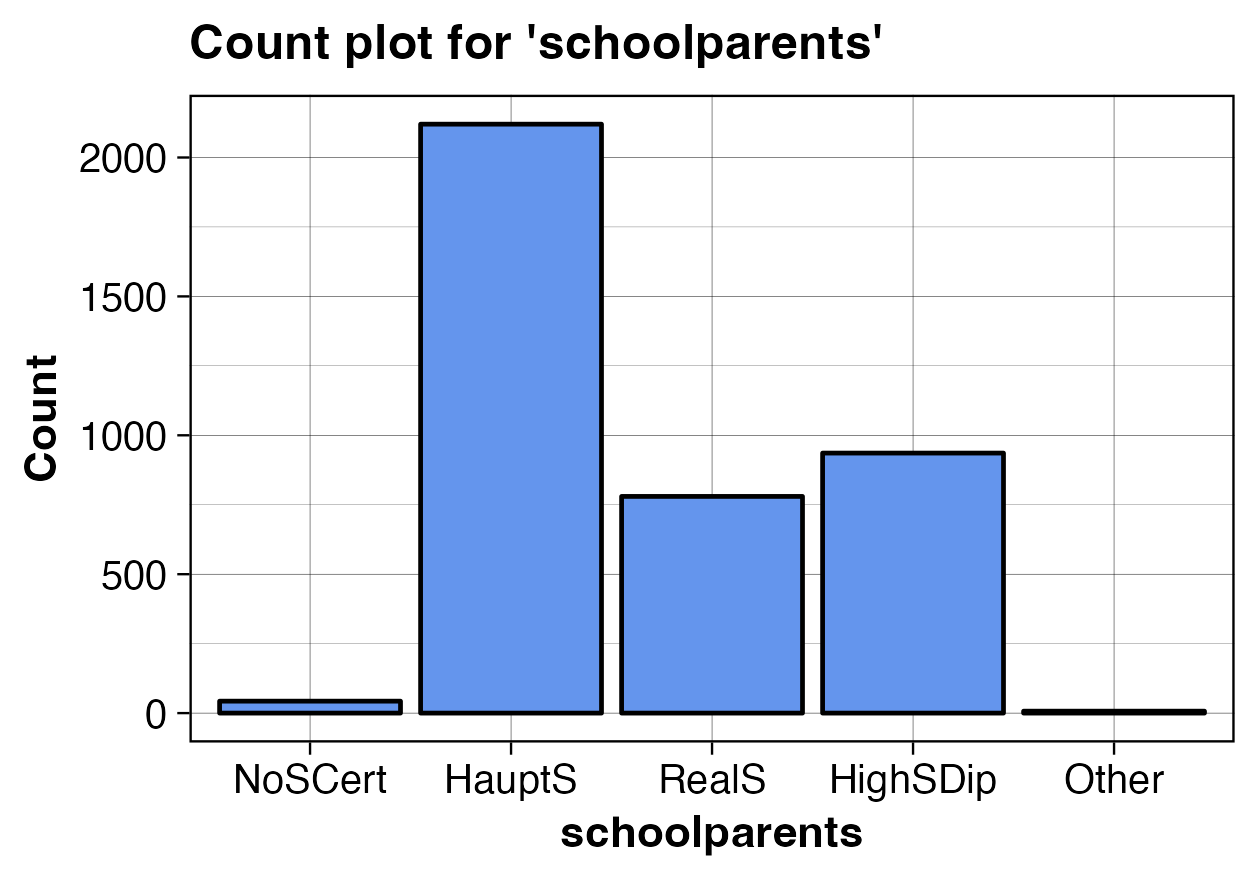} \hfill
		\includegraphics[width=0.475\textwidth]{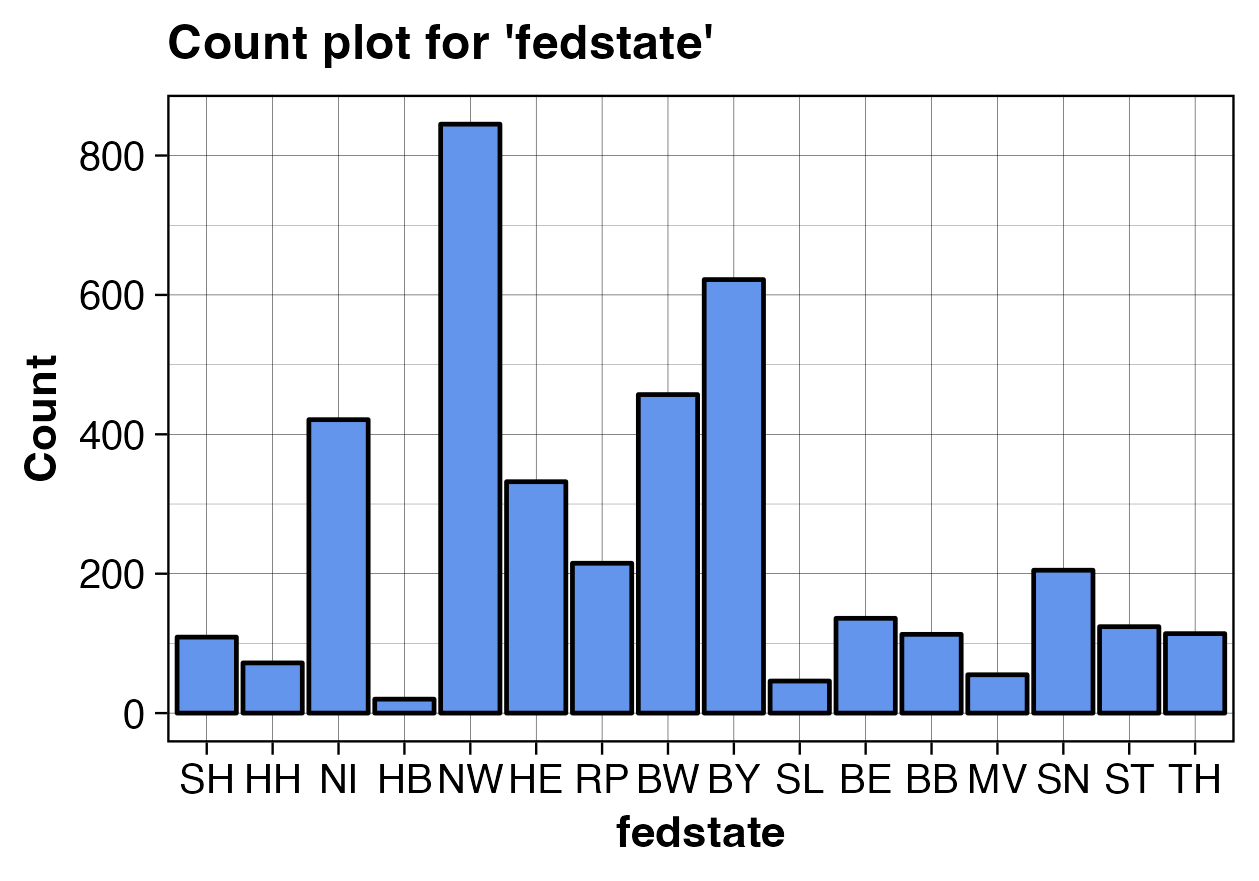} \\
	\end{subfigure}
\end{figure}

\subsection{Target Variable}

\begin{figure}[H]
	\caption{Distribution Plots for the Target Variable}
	\begin{subfigure}{\textwidth}
		\centering
		\includegraphics[width=0.475\textwidth]{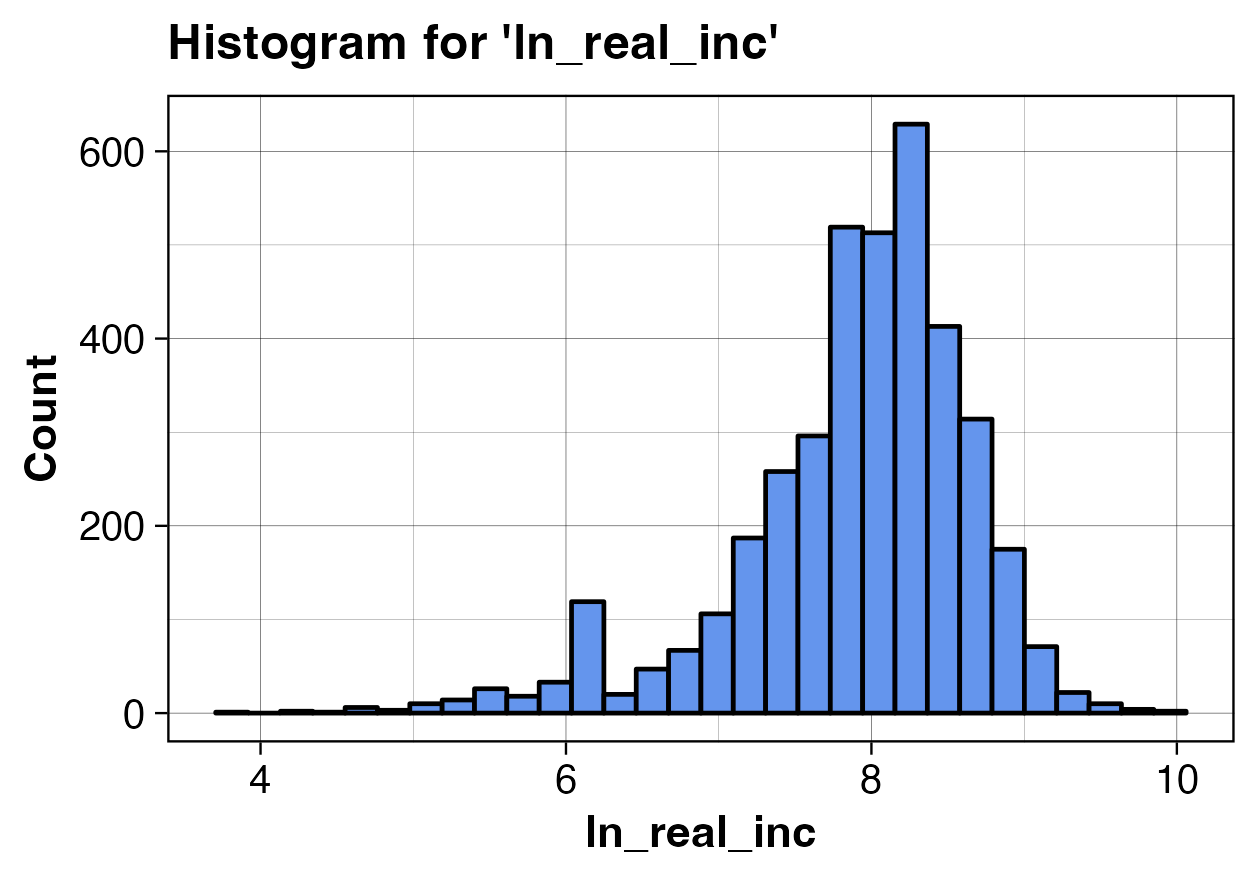} \hfill
	\end{subfigure}
\end{figure}


\newpage
\section{Additional Imputation and Regression Results}

\label{appendix:addResults}

We compare the eight imputation methods on 
three different dimensions: 1. \textbf{Imputation accuracy.} We use two metrics to measure the preservation of true values by the imputation methods, the  Normalized Root Mean Squared Error (NRMSE) and a so-called imputation performance measure (IPM). Here, NRMSE allows for fair comparisons in presence of the different scales \citep{jadhav2019comparison} while IPM 
was proposed by \cite{suh2023comparison} as a summarizing metric for both, numerical and categorical variables, based on the Gower distance.
2. \textbf{Estimation accuracy of correlation coefficients.} Using either Pearson or Spearman correlation coefficients, we compare the resulting correlation matrices of the true and the imputed data sets by means of the Frobenius distance, Mean Absolute Error and RMSE.
3. \textbf{Feature Selection.} We additionally assess the effect of the eight imputation methods on the interpretability of  subsequently applied prediction models. To this end we analyze feature selection after imputation of three common prediction models: An interpretable linear model with LASSO regularization, see, e.g., \citet{hastie2009elements},  and the two tree-based learners Random Forest \citep{breiman2001random} and XGBoost \citep{chen2016xgboost}. While the LASSO approach performs an  automatic feature selection, the two tree-based approaches build their selection on the resulting feature importances \citep{hastie2009elements, chen2016xgboost} by choosing a predefined number of the most important features. In case of multiple imputation, the process has to be adapted as outlined in the Simulation section below. 

\subsection{Imputation Results}

\subsubsection{NRMSE}

\begin{figure}[H]
	\centering
	\caption{NRMSE Results for the chosen Imputation Methods and Missingness Rates}
	\label{fig:NRMSE}
	\includegraphics[width = \textwidth]{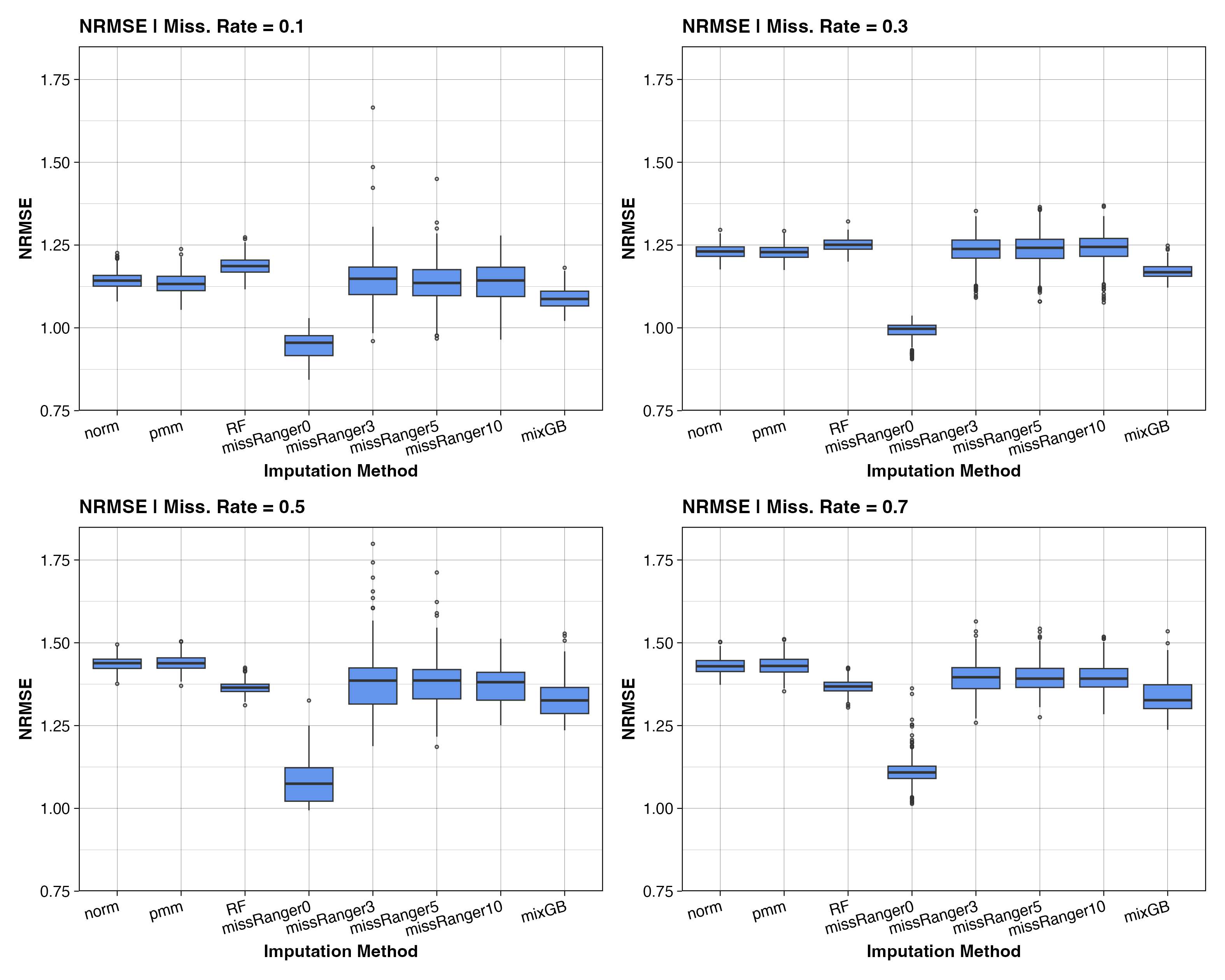}
\end{figure}

The NRMSE results shown in Figure \ref{fig:NRMSE} indicate that \textit{missRanger0} consistently provides the best performance. The \textit{listwise} deletion method is excluded from this analysis because it requires imputed values to compute the distance to the true values, which is not present in this approach.

Looking at the three implementations of \textit{MICE}, both \textit{PMM} and \textit{norm} show similar NRMSE trends across all missingness levels, with NRMSE increasing as missing data rates increase. Compared to Random Forests, these two \textit{MICE} implementations outperform at missingness rates below 30\%, but underperform at rates of 50\% and 70\%. In particular, the NRMSE for Random Forest imputation worsens as the missingness rate increases.

For \textit{missRanger} and \textit{mixGB}, both show a similar pattern across all missingness levels: \textit{missRanger0} leads in performance, followed by \textit{mixGB}, and then by \textit{missRanger} implementations with PMM. There is no clear improvement in NRMSE with respect to the number of donors selected. However, higher missingness rates consistently result in higher NRMSE values, with \textit{missRanger} without PMM maintaining the lowest NRMSE regardless of the missingness rate.

...




\subsubsection{IPM}

\begin{figure}[H]
	\centering
	\caption{IPM Results for each Imputation Method and Missingness Rate}
	\label{fig:IPM}
	\includegraphics[width = \textwidth]{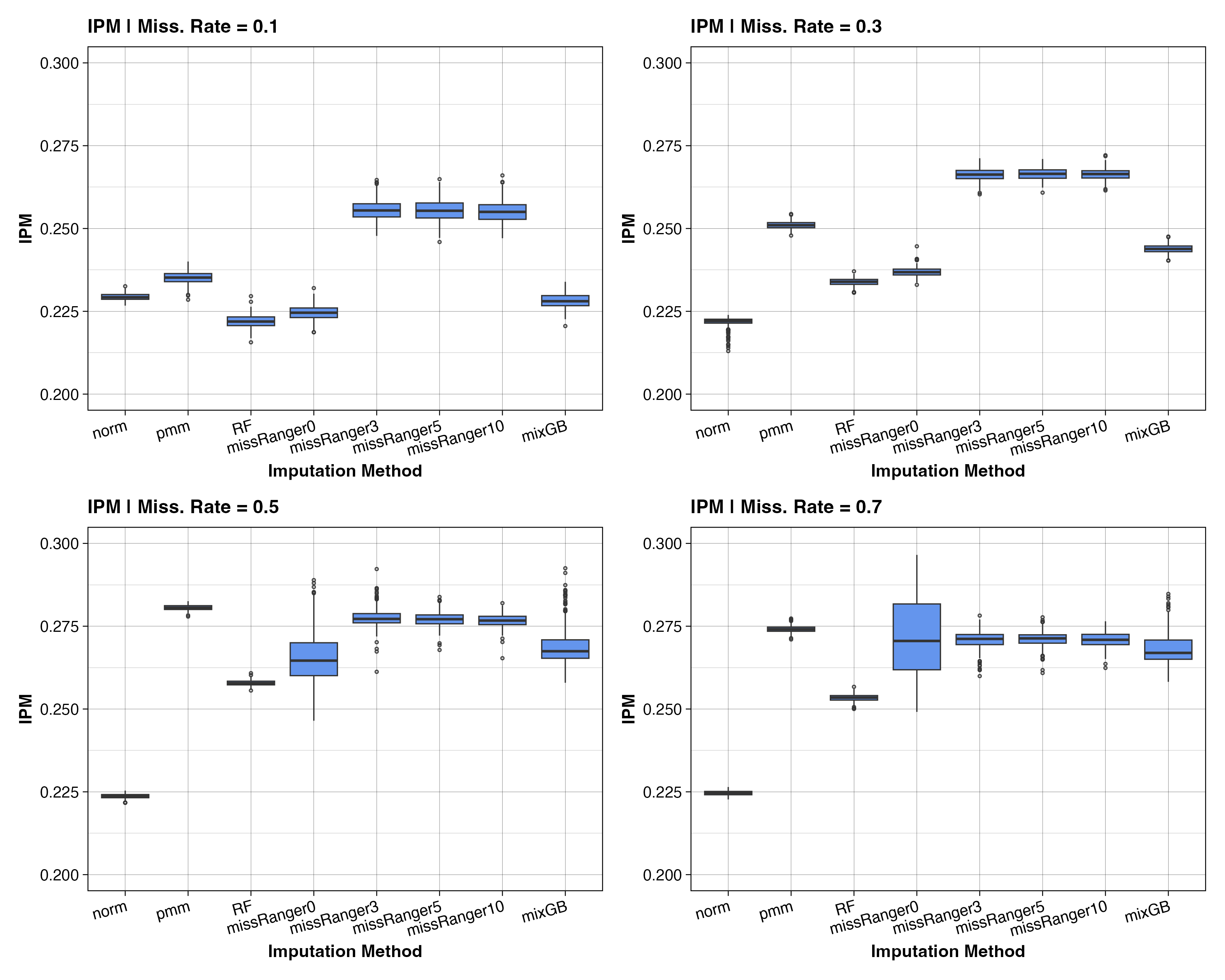}
\end{figure}

Figure \ref{fig:IPM} shows the IPM values for each imputation method over different missingness rates. Notably, the \textit{listwise} deletion method is excluded because IPM measures imputation quality and this method does not perform any imputation, only removes observations with missing data.

Bayesian linear regression (\textit{norm}) ranks first for missingness rates of 30\% and above. Within the \textit{MICE} framework, Random Forests (\textit{RF}) outperforms Predictive Mean Matching (\textit{pmm}) for all missingness rates. Both methods show deteriorating IPM results as missingness increases up to 50\%, but at 70\% IPM decreases again, although it remains higher than at 30\%.

For \textit{missRanger}, PMM results are similar across donor numbers (3, 5, and 10), but consistently worse than \textit{missRanger0} for missingness below 30\%, with the gap narrowing at 70\% missingness. Variance increases for \textit{missRanger} without PMM at 50\% and 70\% missingness, but the number of donors does not significantly affect IPM.
Imputations with \textit{mixGB} show similar median results to \textit{missRanger0}, though slightly worse with lower variance. However, \textit{mixGB} consistently outperforms any \textit{missRanger} implementation with PMM in terms of median IPM.

In summary, excluding \textit{norm} methods, Random Forests in MICE (\textit{RF}) delivers the best IPM results, followed by \textit{missRanger0} and \textit{mixGB} over all missingness rates. At 70\%, \textit{mixGB} surpasses \textit{missRanger} in median performance, although \textit{missRanger} has a wider range of results compared to XGBoost imputations.

\subsubsection{Correlations Review}

Having focused on the predictive accuracy of imputation methods, we now examine how these methods affect data correlations and potential shifts in variable distributions.

\begin{figure}[H]
	\centering
	\caption{Mean Correlation Distances between the Imputed and the Original Dataset, shown by Imputation Method and Missingness Rate}
	\label{fig:corrfull}
	\includegraphics[width = \textwidth]{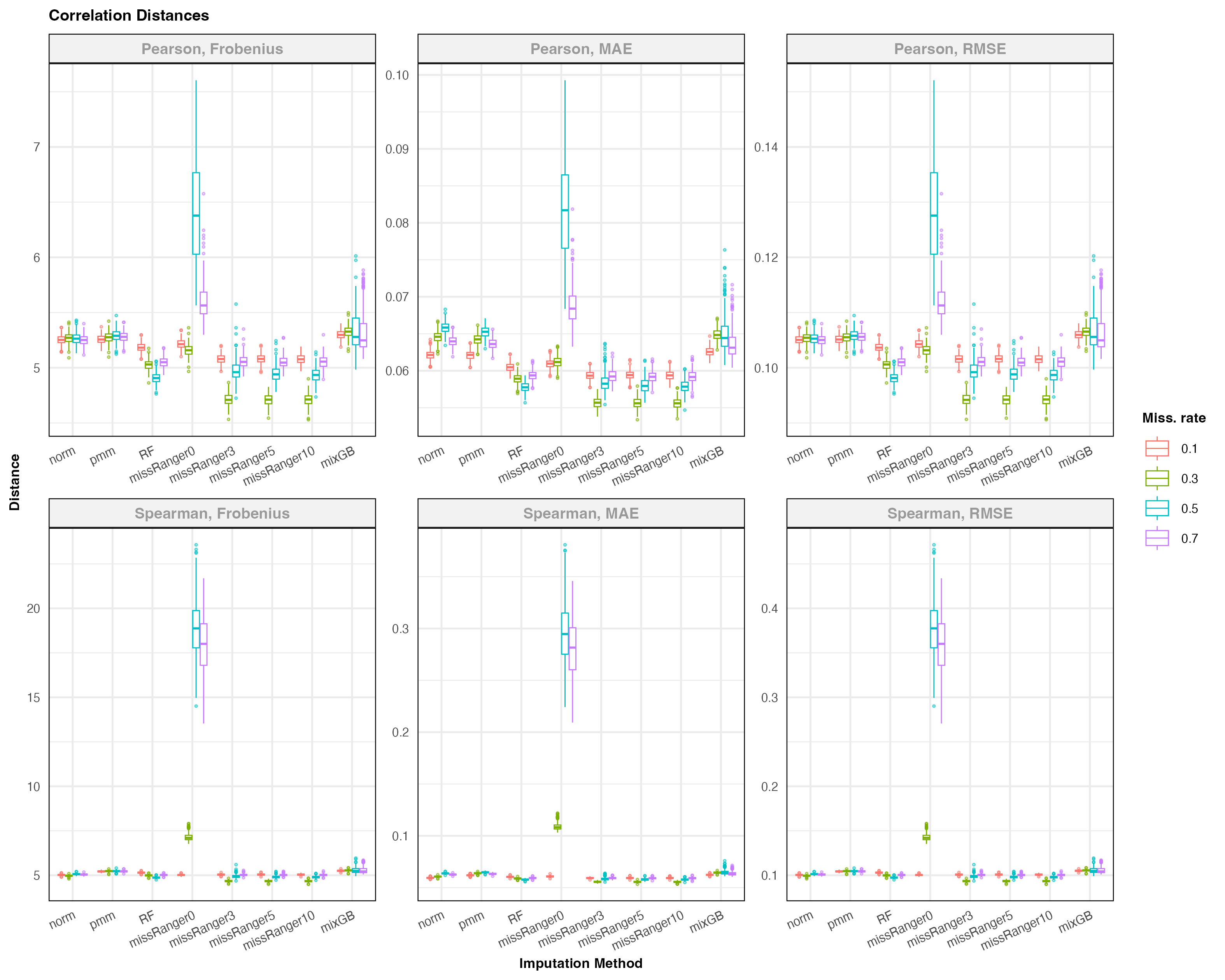}
\end{figure}

Figure \ref{fig:corrfull} illustrates the distances between the correlation matrix of the original dataset (before value amputation) and the correlation matrices of the datasets after imputation. The most significant deviations between metrics are observed for \textit{missRanger0}, especially at missingness rates of 50\% and 70\%. 

While Spearman's coefficients generally indicate similar results, the differences are more pronounced at the 50\% rate compared to the 70\% rate. With the exception of \textit{missRanger} without PMM, most methods show robustness. However, \textit{norm}, \textit{pmm}, and \textit{mixGB} show slightly larger deviations in correlation distances compared to the other methods. 

\subsubsection{Discussion of the Imputation Results}

Table \ref{tab:ImpResults} provides a summary of the imputation method performance across the three metrics NRMSE, IMP and correlation. Methods are categorized based on their effectiveness for each metric: those in the top three are marked with three stars (***), those between fourth and seventh are marked with two stars (**), and the bottom three are marked with one star (*).

For brevity, the results of the NRMSE and IPM have been combined into a single metric, with the calculation details outlined in the Appendix \ref{appendix:normalization}. In this context, \textrm{missRanger0} and \textrm{norm} showed the best performance, followed by \textrm{MICE RF}, \textrm{mixGB}, and \textrm{MICE PMM}.
In terms of correlation distances, the smallest deviations from the baseline case were observed with \textrm{MICE RF}, \textrm{missRanger3}, and \textrm{missRanger5}, followed closely by \textrm{missRanger10}, \textrm{MICE norm}, and \textrm{mixGB}.

Concluding, although \textrm{norm} shows strong performance in terms of (predictive) accuracy, it is important to note, as highlighted by \cite{thurow2021imputing}, that high accuracy in data reproducibility does not necessarily imply that probability distributions are well-preserved. This is evident from the correlation results, where \textrm{norm} does not rank among the top performers. 
Taking this into account, the methods \textrm{RF}, \textrm{missRanger0}, \textrm{mixGB} and \textrm{missRanger10} appear to strike a favorable balance across the two metrics. They provide relatively accurate results within expected ranges and maintain correlation distances better than other methods.

\begin{table}[H]
	\caption{Imputation Results Summary. According to each metric the top 3 get 3 stars (***), the next 3 get 2 stars (**) and the lowest 3 get 1 star (*).} \label{tab:ImpResults}
	\centering
	\small
	\begin{tabulary}{\textwidth}{lccs}  
		\toprule[2 pt]
		\textbf{Method} & \textbf{Accuracy} & \textbf{Correlation} & \textbf{Time} \\
		\midrule
		\textrm{listwise}  & * & * & ***  \\ 
		\textrm{MICE norm}  & *** & ** & *** \\ 
		\textrm{MICE PMM}  & **  & * & **\\ 
		\textrm{MICE RF}  & *** & *** & ** \\ 
		\textrm{missRanger.0}  & ***  & * & *\\ 
		\textrm{missRanger.3}  & * & *** & ** \\ 
		\textrm{missRanger.5}  & * & *** & * \\ 
		\textrm{missRanger.10}  & ** & ** & * \\ 
		\textrm{mixGB}  & ** & ** & ** \\ 
		\bottomrule[2 pt]
	\end{tabulary}
\end{table}

\subsection{Regression}
\subsubsection{Baseline Regression Results}

Table \ref{tab:MSEbase} provides the summary statistics from 100 iterations used to train the three statistical learners models for the complete dataset as a baseline or best case scenario. 
From these results, XGBoost (\texttt{regr.xgboost}) achieves the lowest MSE minimum, median, mean and maximum, closely followed by Random Forests (\texttt{regr.ranger}), and then LASSO (\texttt{regr.cv\_glmnet}). I.e. the outcome variable is best explained using XGBoost when using the complete data. 

\begin{table}[H]
	\caption{Summary Statistics for the MSE computed in the Complete Dataset.} \label{tab:MSEbase}
	\centering
	\small
	\begin{tabulary}{\textwidth}{lccccccc}  
		\toprule[2 pt]
		\textbf{ML Method} & \textbf{Min} & \textbf{1st. Qu.} & \textbf{Median} & \textbf{Mean} & \textbf{3rd. Qu.} & \textbf{Max} & \textbf{SD}\\
		\midrule
		\textrm{LASSO} & 0.182 & 0.185 & \textbf{0.186} & 0.186 & 0.187 & 0.189 & 0.002 \\ 
		\textrm{Random Forest} & 0.138 & 0.141 & \textbf{0.142} & 0.142 & 0.143 & 0.146 & 0.002 \\
		\textrm{XGBoost} & 0.132 & 0.134 & \textbf{0.135} & 0.136 & 0.137 & 0.140 & 0.002 \\ 
		\bottomrule[2 pt]
	\end{tabulary}
\end{table}



\subsubsection{Regression Results in Comparison}

\begin{figure}[H]
	\centering
	\caption{Mean Squared Error for each Imputation Method. The horizontal Lines are the Median of the MSEs obtained for the Complete Dataset}
	\label{fig:MSE}
	\includegraphics[width = \textwidth]{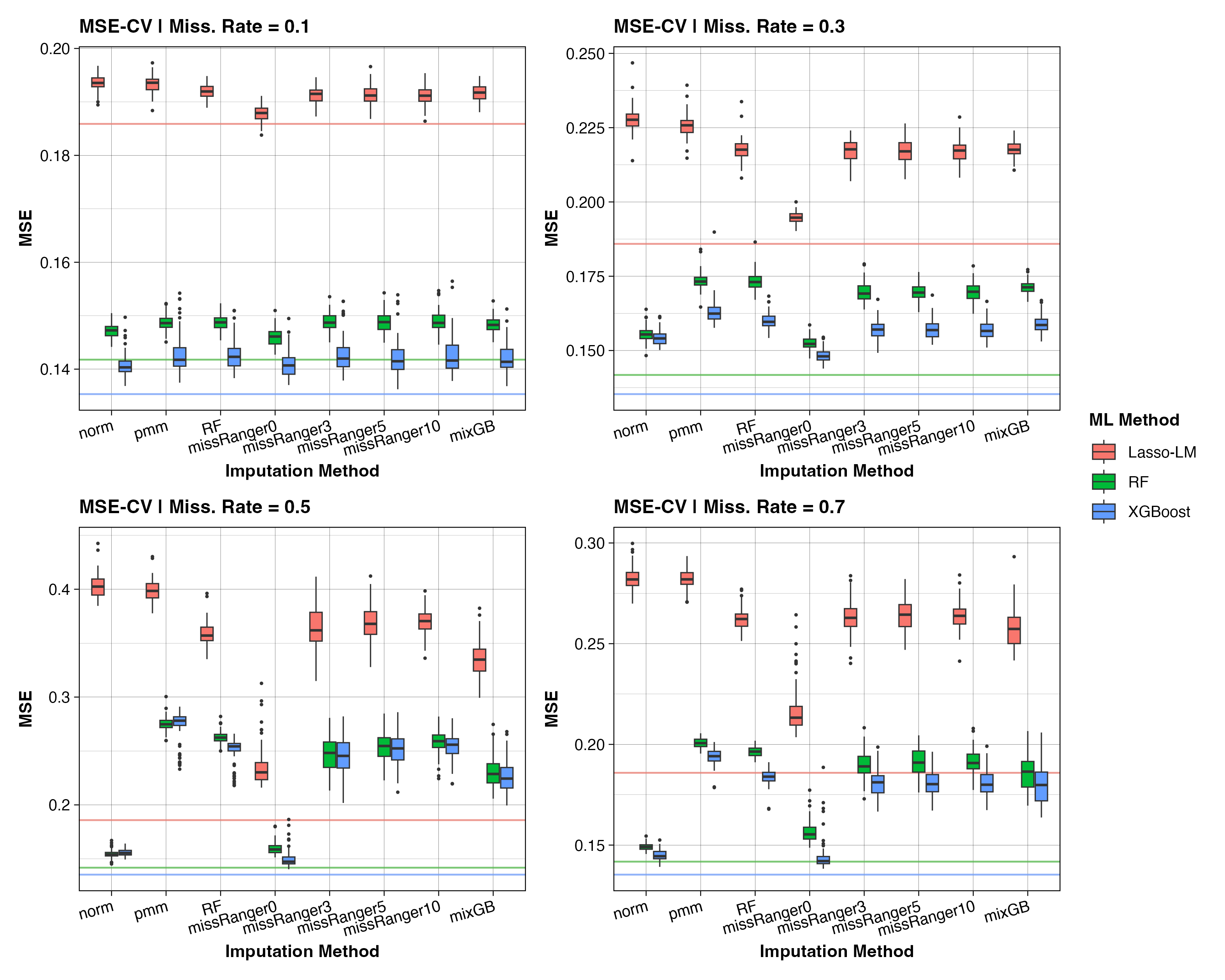}
\end{figure}

Figure \ref{fig:MSE} shows the Mean Squared Errors (MSEs) for each combination of statistical learner and imputation method for each of the  missingness rates per panel. In addition, the horizontal lines indicate the median MSE using the complete data, as detailed in Table \ref{tab:MSEbase}. The color coding of these lines is consistent with the legend.

The results for listwise deletion are omitted from Figure \ref{fig:MSE} to improve the overview, because combination of the statistical learners and listwise deletion resulted in outliers making it difficult to detect smaller differences between the other combinations\footnote{Particularly for the highest missingness rates 50\% and 70\%, listwise deletion reduces the size of the dataset to the point where there is insufficient data to train valid models under the nested resampling framework employed. In some cases, datasets were completely depleted of entries. In addition, training the LASSO model proved at times infeasible in certain scenarios due to features with singular observations (such as subsets containing only one gender), which resulted in computational failures. In addition, the MSE values for scenarios where simulations yielded results ranged from 0.5 to 1.25, making the box plots for other methods too small for meaningful analysis.}. Results incorporating listwise deletion can be found in  Appendix \ref{appendix: Plots}.

When evaluating the machine learning models, LASSO consistently has the highest error relative to the other models under identical missingness rates and imputation methods, which is in-line with the baseline using complete data. XGBoost consistently achieves the lowest MSE, except at a 50\% missingness rate for the MICE \textrm{norm} and \textrm{pmm} imputations, where the Random Forest models show a slightly lower median MSE.

The influence of increasing missingness rates generally leads to increased MSEs, peaking at 50\% before declining slightly at 70\%, though not to the levels observed at 30\%. Notably, the \texttt{norm}, and \texttt{missRanger0} imputation methods show a less pronounced increase in MSE with increasing missingness, suggesting improved robustness. These three methods also consistently outperform other imputation techniques in terms of MSE across the range of missingness rates examined. In all contexts, the model predictions fail to exceed the median MSE achieved with the complete dataset. The only exception is that the minimum of the combination of LASSO with missRanger without PMM at 10\% missingness was lower than the median baseline.

Examining \textrm{missRanger}, the inclusion of PMM appears to increase the MSE compared to its counterpart without PMM. The number of donors (3, 5, and 10) does not significantly change the MSE results when PMM is applied. Tree-based imputation methods generally perform similarly across MSE assessments at all missingness rates, with \textrm{mixGB} showing slightly better results at missingness rates of 50\%. 

\subsection{Discussion of MSE results}

In terms of MSE results, XGBoost shows superior performance, especially with non-PMM methods such as \textrm{missRanger0} and \textrm{norm} and a missingness rate of 10\%, 30\% and 70\%. Only for 50\% are the MSE results for XGBoost and Random Forest mostly overlapping. Regardless of the statistical learner, \textrm{listwise} consistently produces the worst results. In addition, increasing the missingness rate generally increases the MSE up to 50\%, followed by a slight decrease at 70\%, but not enough to reach the levels observed at 30\%. There is also no evidence that using the same methods for imputation and regression (e.g., \textrm{mixGB} and XGBoost) is better than using different methods for imputation than for regression (e.g., \textrm{MICE PMM} and XGBoost).

For \textrm{missRanger}, including the number of donors \texttt{pmm.k} tends to increase the MSE, and there is no significant improvement or reduction in MSE between values 3, 5, and 10. Using \textrm{mixGB} for imputation produces very similar (missingness rate = 10\% or 30\%) or slightly better results than  \textrm{missRanger} with PMM (missingness rate = 50\% or 70\%).

\newpage
\section{Additional MSE Results}
\label{appendix: Plots}
\subsection{MSE including listwise deletion}

\begin{figure}[H]
	\centering
	\caption{Mean Squared Error for each Imputation Method including listwise deletion. The horizontal Lines are the Median of the MSEs obtained for the Complete Dataset}
	\label{fig:MSE_listwise}
	\includegraphics[width = \textwidth]{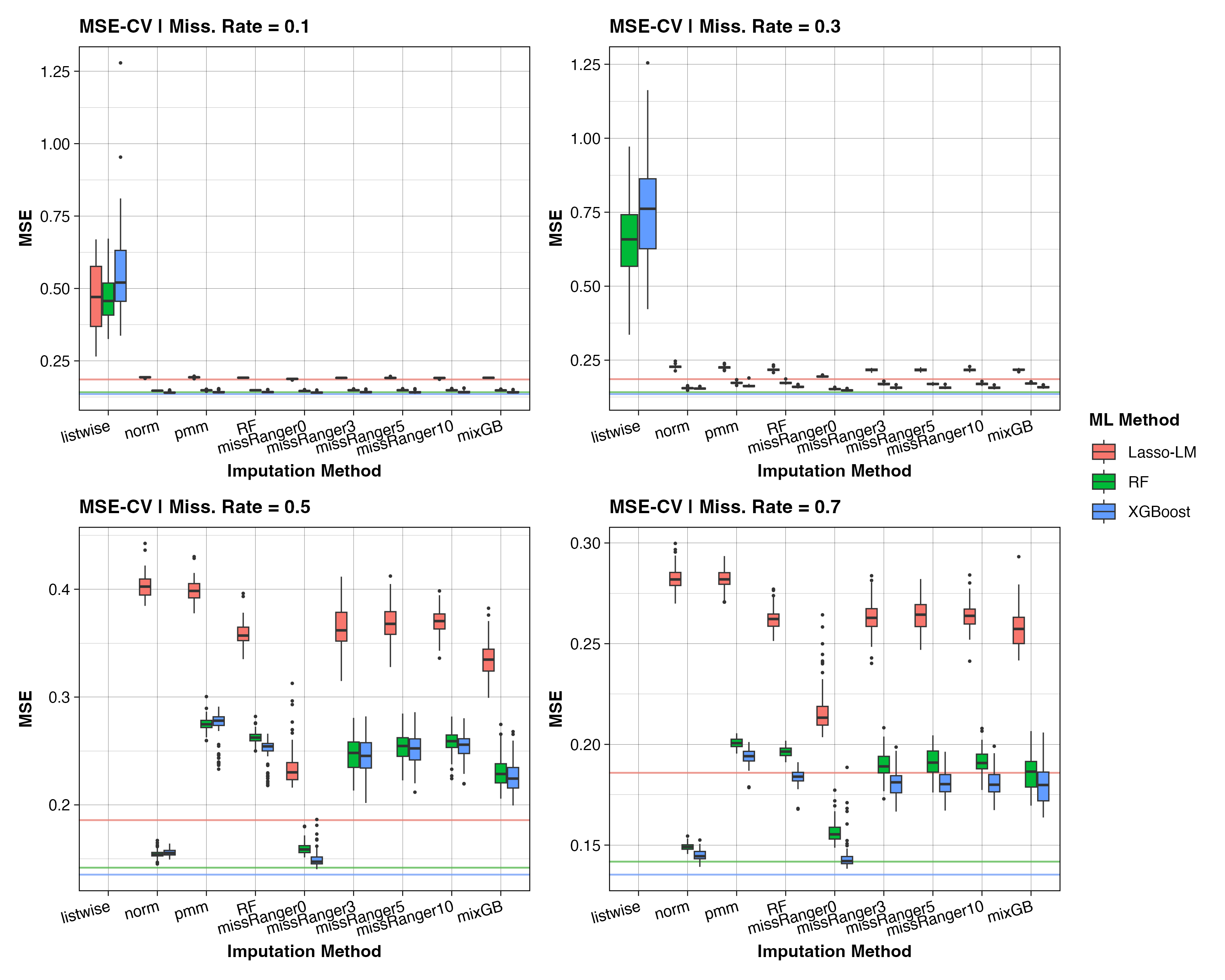}
\end{figure}

\newpage
\section{Imputation Accuracy Normalization Details}
\label{appendix:normalization}

Table \ref{tab:Normalization} has the results of the normalizations for both NRMSE and IPM into one single average. The second and the third column show the median of the metrics for each imputation method (at all missingness rates). Then, these values are normalized each to a scale between 0 and 1, being 0 the value for the lowest calculation and 1 for the maximum. And then these two normalized values are averaged per method. 

\begin{table}[H]
	\caption{Detailed Calculations for the normalization of the NRMSE and IPM} \label{tab:Normalization}
	\centering
	\footnotesize
	\begin{tabular}{lcccc|c}
		\hline
		\textbf{Method} & \textbf{NRMSE (median)} & \textbf{IPM (median)} & \textbf{Norm. NRMSE} & \textbf{Norm. IPM} & \textbf{Average} \\ 
		\hline
		\textrm{missRanger0} & 1.034 & 0.251 &  0.000    &  0.548  & \textbf{0.274} \\ 
		\textrm{norm} & 1.330 & 0.225 &  0.909  &  0.000  & \textbf{0.455} \\ 
		\textrm{RF} & 1.305 & 0.244 &  0.832  &  0.393  & \textbf{0.613} \\ 
		\textrm{mixGB} & 1.250 & 0.256 &  0.664  &  0.640  & \textbf{0.652} \\ 
		\textrm{pmm} & 1.330 & 0.263 &  0.909  &  0.785  & \textbf{0.847} \\ 
		\textrm{missRanger10} & 1.310 & 0.269 &  0.848  &  0.909  & \textbf{0.878} \\ 
		\textrm{missRanger5} & 1.315 & 0.269 &  0.863  &  0.909  & \textbf{0.886} \\ 
		\textrm{missRanger3} & 1.315 & 0.269 &  0.863  &  0.909  & \textbf{0.886} \\ 
		\textrm{listwise} & \underline{1.360} & \underline{0.273} &  1.0000  &  1.0000  & \textbf{1.000} \\ 
		\hline
	\end{tabular}
\end{table}

For the \textrm{listwise} deletion case, given that there was no computation, both the NRMSE and IPM were assigned to be equal to the maximum observed value per case plus 10\% of the difference between the maximum and the minimum (the range) observed values per metric. These values are shown with \underline{underlined} numbers.

\subsection{Discussion}

With regard to the performance of the imputation models, the results indicated that missRanger.0 and MICE norm exhibited promising outcomes with respect to accuracy, as evidenced by the NRMSE and IPM metrics, which are measures of reproducibility. However, the results indicated that there were significant discrepancies between the imputed and original datasets, as evidenced by the significant correlation distances observed between the two.

The results of imputation models indicate that MICE with PMM and RF, along with missRanger.3 and mixGB, offer an optimal balance between accuracy and minimal correlation variance, while maintaining efficient computational times (not shown in the manuscript).

With regard to the MSE results obtained subsequent to imputation, XGBoost demonstrated superior performance relative to LASSO and Random Forests under identical missingness rates and imputation methods. Specifically, at missingness rates exceeding 30\%, XGBoost in conjunction with missRanger.0 and MICE norm exhibited lower MSEs than other imputation methods.

With regard to missRanger, the study revealed that in the absence of PMM, it demonstrated consistent and robust imputation accuracy (NRMSE and IPM) and prediction error (MSE) across a range of missingness rates. However, it was observed that the correlation distances were relatively larger, particularly above 30\%. The incorporation of PMM resulted in enhanced correlation distances, yet it led to a deterioration in NRMSE, IPM, and MSE. No notable alterations were observed when the number of donors was varied. 



\end{document}